\documentclass{sig-alternate}
\usepackage{algorithmicx}
\usepackage{algorithm}
\usepackage{algcompatible}
 \usepackage{algpseudocode}
\usepackage{graphicx}   \usepackage{balance}    \usepackage{subfigure}
\usepackage{multirow}
\usepackage{pbox}
\usepackage{cite}
\usepackage{float}
\usepackage[table]{xcolor}\usepackage{epstopdf}
\usepackage{amsmath}
\usepackage{hhline}
\usepackage{rotating}
\usepackage{booktabs}
\usepackage{xparse}
\usepackage[font={small}]{caption}
\def \sys {\textit{semMatch}}
\NewDocumentCommand{\rot}{O{45} O{0.4em} m}{\makebox[#2][l]{\rotatebox{#1}{#3}}}
\NewDocumentCommand{\ret}{O{45} O{2.4em} m}{\makebox[#2][l]{\rotatebox{#1}{#3}}}

\newcommand{\rulesep}{\unskip\ \vrule\ }

\newtoks\copyrtyr
\newtoks\acmcopyr
\newtoks\boilerplate
\newtoks\acmdoi
\global\acmcopyr={978-1-4503-2521-9/13/11}  
\global\copyrtyr={2013}

\global\boilerplate={
\vspace{5mm}
Permission to make digital or hard copies of all or part of this work for personal or classroom use is granted without fee provided that copies are not made or distributed for profit or commercial advantage and that copies bear this notice and the full citation on the first page. Copyrights for components of this work owned by others than ACM must be honored. Abstracting with credit is permitted. To copy otherwise, or republish, to post on servers or to redistribute to lists, requires prior specific permission and/or a fee. Request permissions from Permissions@acm.org.\\
SIGSPATIAL'15, November 03-06, 2015, Bellevue, WA, USA \\
{\small\copyright} 2015 ACM. ISBN 978-1-4503-3967-4/15/11...\$15.00\\
DOI: http://dx.doi.org/10.1145/2820783.2820824
}
\global\acmdoi= {}\newtoks\copyrightetc
\global\copyrightetc{}\toappear{\the\boilerplate\par
{\confname{\the\conf}} \the\confinfo\par \the\copyrightetc\par \the\acmdoi.}

\makeatletter

\begin{document}

\title{semMatch: Road Semantics-based Accurate Map Matching for Challenging Positioning Data}

\numberofauthors{2}
\author{
  \alignauthor Heba Aly\\Department of Computer Science\\
  University of Maryland, USA
  \email{heba@cs.umd.edu}
    \alignauthor Moustafa Youssef\thanks{Moustafa Youssef is currently on sabbatical from Alexandria University, Egypt.}\\   Wireless Research Center\\
 E-JUST, Egypt
  \email{moustafa.youssef@ejust.edu.eg}
  }
\maketitle
\begin{abstract}
Map matching has been used to reduce the noisiness of the location estimates by aligning them to the road network on a digital map. A growing number of applications, e.g. energy-efficient localization and cellular provider side localization, depend on the availability of only sparse and coarse-grained positioning data; leading to a challenging map matching process.

In this paper, we present \sys{}: a system that can provide accurate HMM-based map matching for challenging positioning traces. \sys{} leverages the smartphone's inertial sensors to detect different road semantics; such as speed bumps, tunnels, and turns; and uses them in a mathe-matically-principled way as hints to overcome the sparse, noisy, and coarse-grained input positioning data, improving the HMM map matching accuracy and efficiency. To do that, \sys{} applies a series of preprocessing modules to handle the noisy locations. The filtered location data is then processed by the core of \sys{} system using a novel incremental HMM algorithm that combines a semantics-enriched digital map and the car's ambient road semantics in its estimation process.

We have evaluated \sys{} using traces collected from different cities covering more than 150km under different harsh scenarios including coarse-grained cellular-based positioning data, sparse GPS traces with extremely low sampling rate, and noisy traces with a large number of back-and-force transitions. The results show that \sys{} significantly outperforms traditional map matching algorithms under all scenarios, with an enhancement of at least 416\% and 894\% in precision and recall respectively in the most difficult cases. This highlights its promise as a next generation map matching algorithm for challenging environments.

\end{abstract}



\section{Introduction}\label{sec:intro}
With location-based services (LBS) becoming an integral part of our everyday life, map matching; the problem of finding which road the vehicle is on given noisy input location traces; has gained attention with a large number of applications including car navigation, enriching map semantics, traffic estimation, among others. A number of map matching techniques have been proposed in literature \cite{greenfeld2002matching,quddus2003general,yang2005map,li2007practical,alt2003matching,brakatsoulas2005map,hummel2006map,newson2009hidden,wang2014eddy}, mainly designed for GPS positioning data due to its worldwide availability and relatively high accuracy.
However, there are a number of situations where coarse-grained network-based location information is the only viable solution, e.g. in situations requiring energy-efficient localization, localization from the cellular provider side, and/or localization using devices that do not include a GPS chip~\cite{cellsense,gac,cellsense2,ibrahim2011hidden}. Therefore, a number of map matching techniques started to emerge that leverage other sensors on the phone (e.g. WiFi~\cite{vtrack}, cellular fingerprinting \cite{ctrack} with neighbouring cell tower information\footnote{The majority of cell phones in the market only give access to the associated cell tower information only with no access to the neighbouring cell tower information.}, and inertial sensors~\cite{autowitness,wheelloc})  to enhance the cellular-based localization accuracy and hence provide better map matching accuracy. Nevertheless, the sensors used are still noisy and/or consume high power, leading to coarse-grained localization; and hence lower map matching accuracy.

\begin{figure*}[!t]
\centering
\subfigure[Cellular pos. data suffer from back-and-forth trans. The loc. estimates seq. (C1-C2-C3-C2-C4) bounces bet. C2 and C3 due to the overlapping coverage bet. the cell towers in the shaded Area A.]{
      \includegraphics[width=0.31\linewidth]{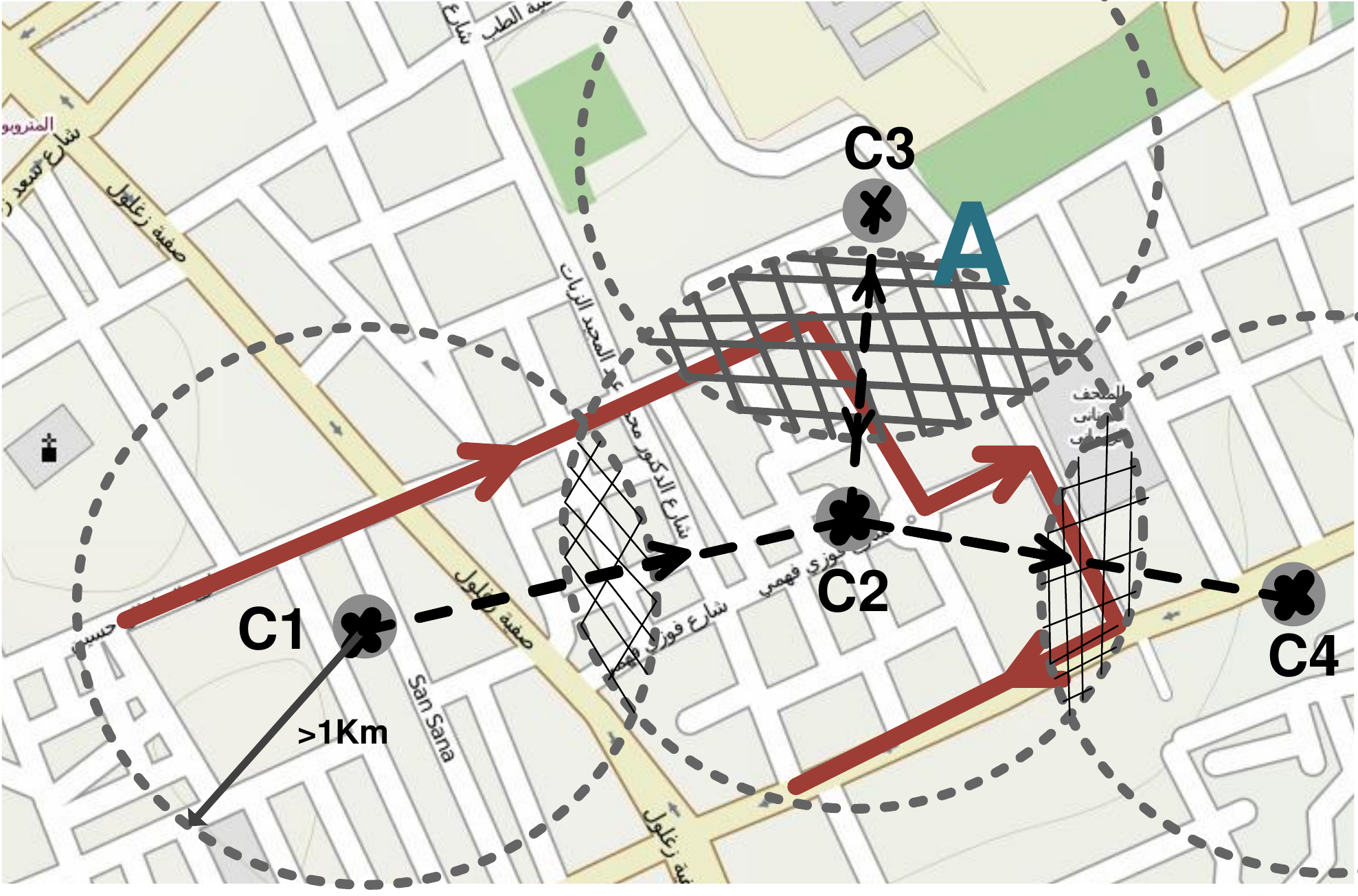}
      \label{fig:ch_pong}
    }
    \subfigure[Cellular positioning data are sparse with errors in the order of kilometers. The user driving path is fully covered by 2 cell towers (i.e. 2 location estimates).]{
      \includegraphics[width=0.31\linewidth]{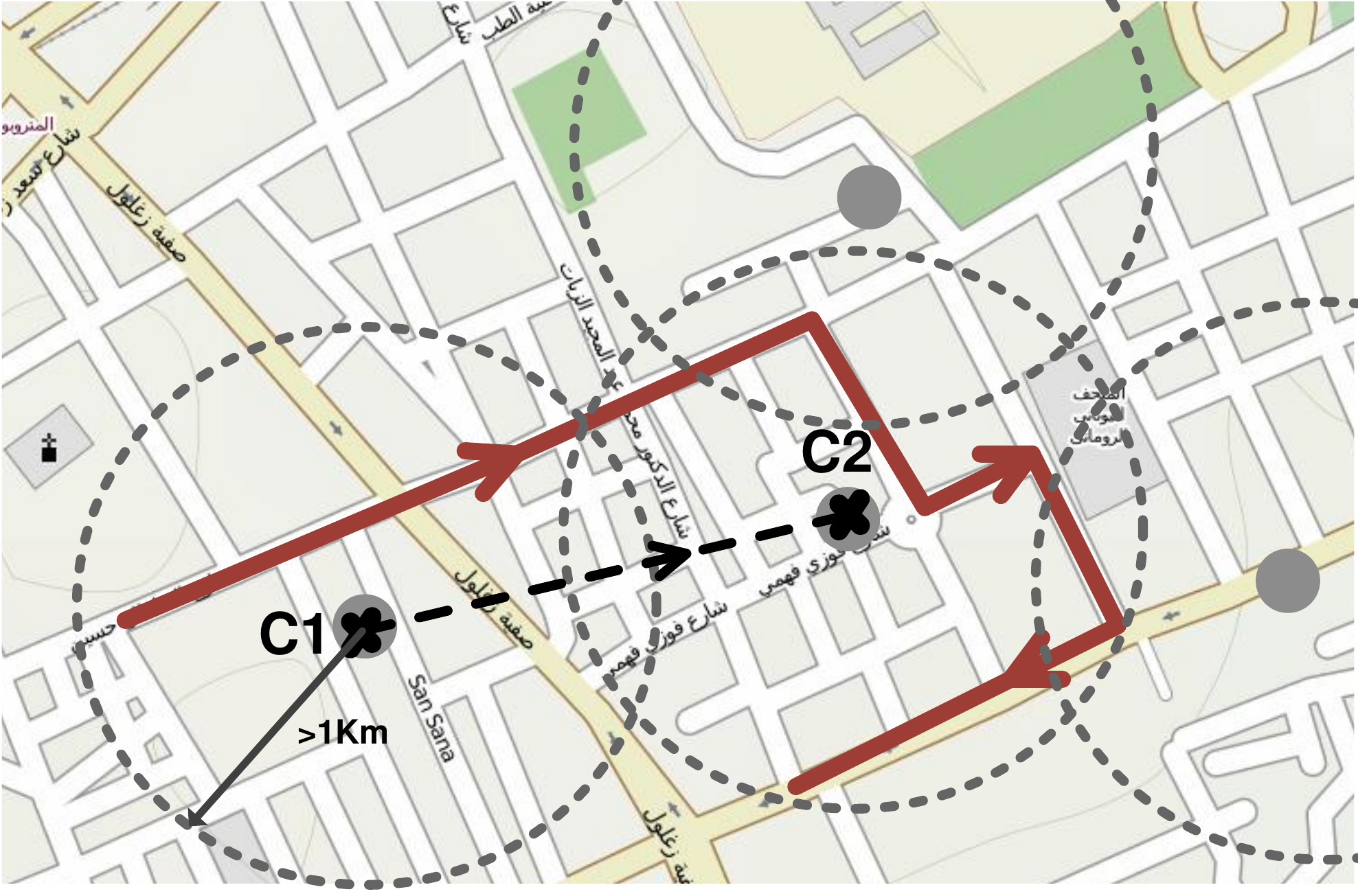}
      \label{fig:ch_sparse}
    }
    \subfigure[Lots of possible map matching routes and the actual route may be difficult to deduce. For example, routes 1 and 2 represent more probable map matching output by previous approaches, e.g.~\cite{newson2009hidden,mohamed2014accurate}.]{
      \includegraphics[width=0.31\linewidth]{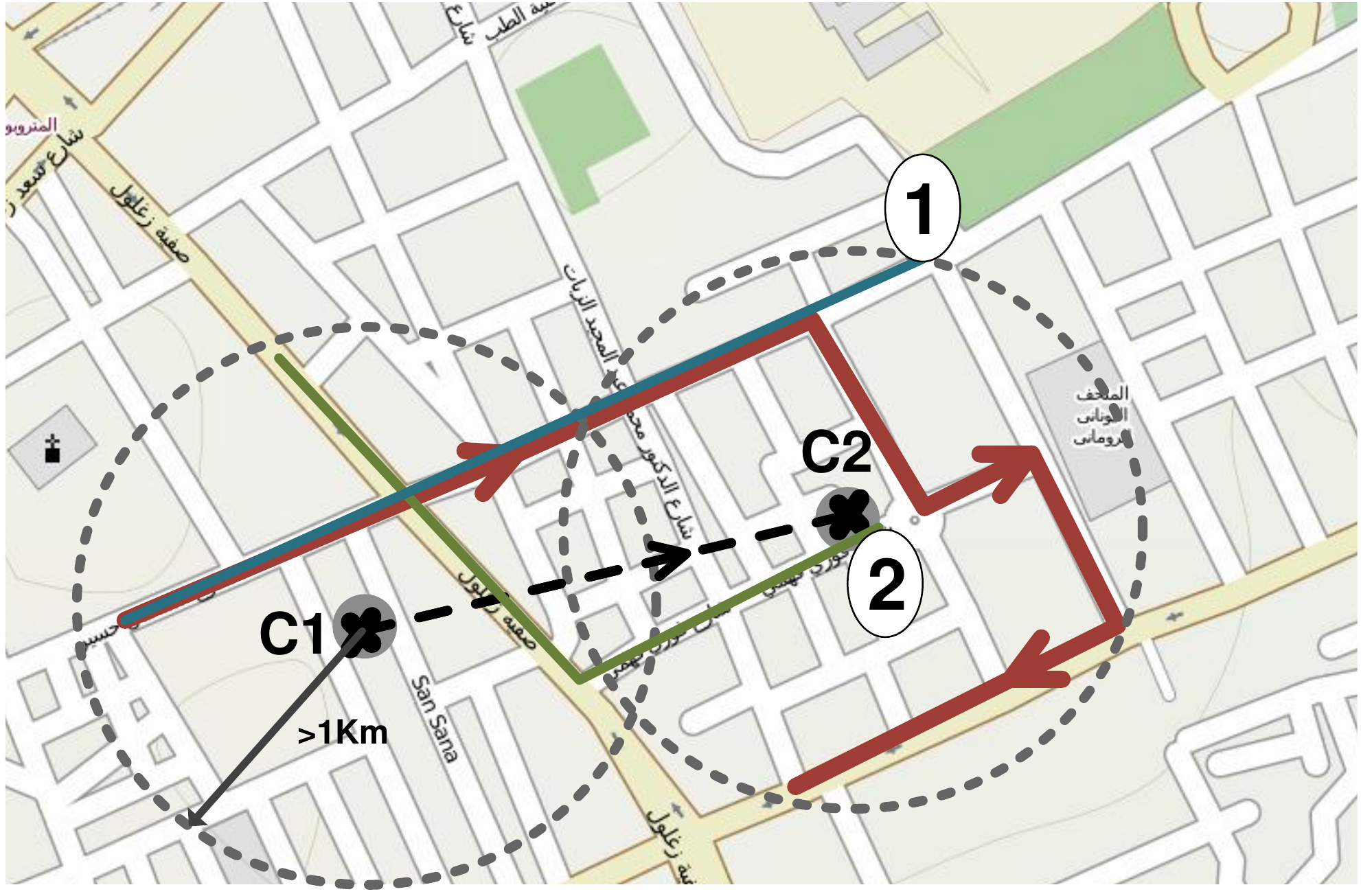}
      \label{fig:ch_map_ex}
    }
\caption{Illustration of the cellular-based map matching challenges. The car moves on the red solid line. The cellular tower position is the circle and its coverage is highlighted with the dotted circle.}
\label{fig:cell_challenges}
\end{figure*}

\begin{figure}[!t]
\centering
\includegraphics[width=0.7\linewidth]{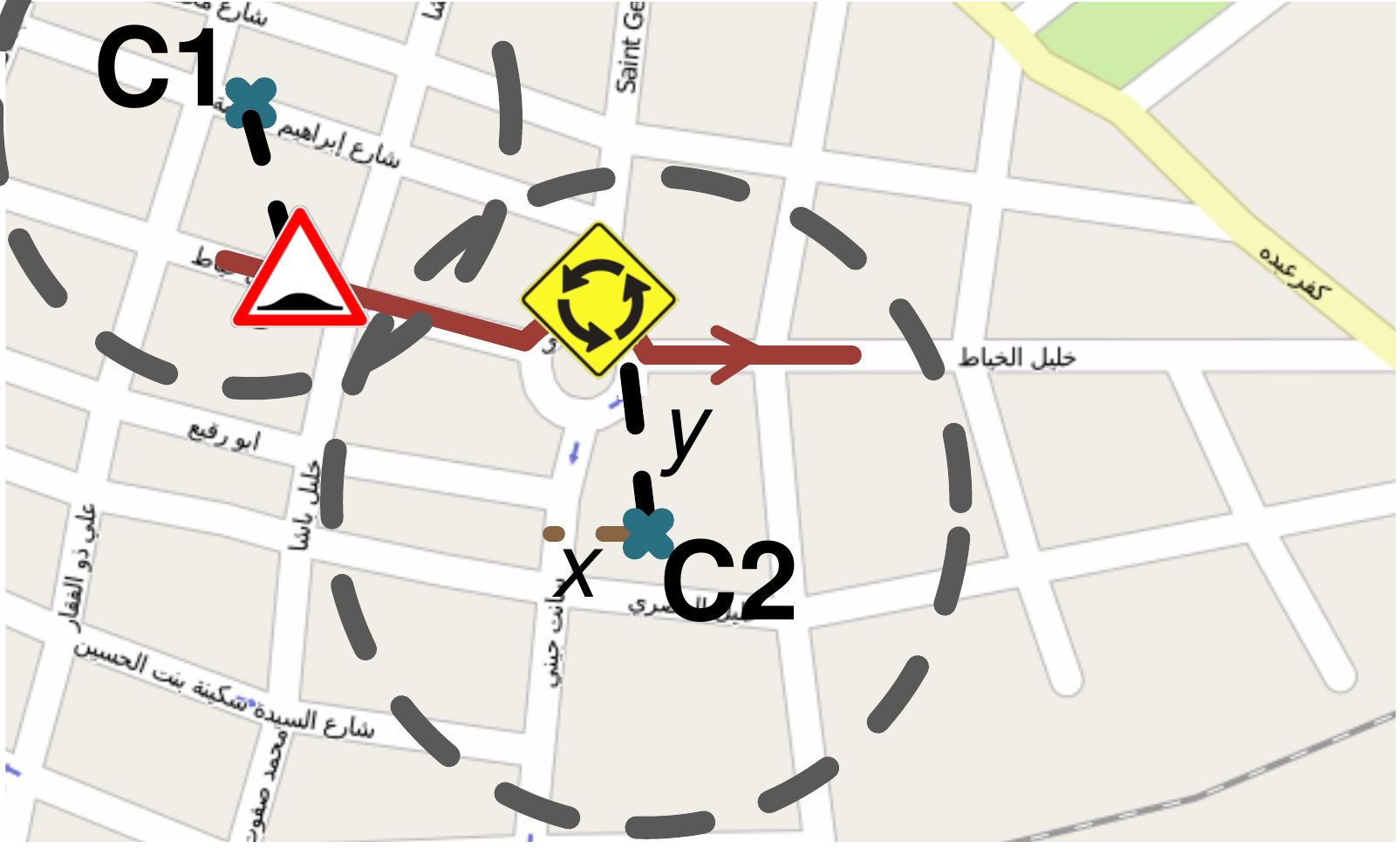}
\caption{Traditional map matching techniques favor roads that are closer to the input location estimates. With coarse-grained accuracy, the  input locations, e.g. C1 and C2, are far from the correct user trajectory. Knowledge of surrounding road semantics sensed by the energy-efficient phone sensors can reduce the ambiguity of the roads, leading to reduced complexity and high map matching accuracy.}
\label{fig:sem_mm_ex}
\end{figure}

In this paper, we target map matching for \textbf{\emph{challenging positioning data that is characterized by sparse, noisy, and/or coarse-grained input location estimates}}. For example, cell ID-based cellular-localization systems; which estimate the user location as her associated cell tower location; are ubiquitous and energy-efficient since cellular data is available on all cell phones and using it for localization does not consume extra energy in addition to the normal phone operation as opposed to the GPS or WiFi-based systems. However, for such positioning data, the typical assumptions of traditional map matching algorithms do not hold (Figure~\ref{fig:cell_challenges}). Specifically, the location error is in the order of kilometers and the input location samples are highly sparse (Figure~\ref{fig:ch_sparse}). Moreover, the location estimates suffer from back-and-forth transitions (aka ``Ping-Pong'' effect~\cite{pingpong1,pingpong2}) due to the handover between different cell-towers for call quality maintenance (even with no movement from the user) (Figure~\ref{fig:ch_pong}).  These stringent characteristics lead to a much harder map matching problem in terms of low quality of the input location data and the high number of candidate road segments (Figure~\ref{fig:ch_map_ex}); affecting both the accuracy and computational complexity.

We therefore present \sys{}, a system for accurate and efficient map matching of challenging positioning data. \sys{} leverages the commonly available and energy-efficient inertial sensors to detect the car's ambient road semantics, such as speed bumps and tunnels, and uses them as hints to guide the map matching process and infer the car current road segment (Figure~\ref{fig:sem_mm_ex}). At the core of \sys{} is a \emph{novel} incremental Hidden Markov Model (HMM) algorithm that takes the car's road surrounding information and the noise of the input data into account to enhance the accuracy of the estimated road segments and efficiently handle the increased number of candidate road segments. This HMM is combined with a number of preprocessing modules that reduce the noise in the input data.

We have implemented \sys{} and evaluated it on driving traces collected in multiple cities. Our results show that \sys{} can achieve a map matching F-measure of up to 97\% under different challenging scenarios including coarse-grained cellular location information, highly sparse GPS traces, and traces with a lot of back-and-force transitions. Moreover, its performance is significantly better than traditional map matching techniques~\cite{newson2009hidden,mohamed2014accurate} under all scenarios, reaching an enhancement of at least (416\%, 894\%, 232\%) in (precision, recall, F-measure) respectively in the most challenging cases.

In summary, our main contributions in this paper are four-fold:
\begin{itemize}
\item We present the architecture of \sys{}: a real-time map matcher for challenging positioning data. The system can accurately and efficiently map match traces with accuracy in the order of kilometers, traces with low update rates (e.g. one update every two minutes), and noisy traces with back-and-force transitions.

\item We present a number of preprocessing modules that reduce the noise in the phone sensors and input locations.

\item We discuss the details of a novel HMM framework that integrates a digital map with the car's ambient road semantics knowledge to accurately detect the car road segment and reduces the computational complexity.

\item We implement the \sys{} system and evaluate its performance using real driving
traces covering more than 150 Km and compare it to the state-of-the-art HMM-based map matching algorithms under different harsh conditions.
\end{itemize}

The rest of the paper is organized as follows: Section~\ref{sec:sys_ov} gives an overview of the \sys{} system and its different components. Then, we give the details of the proposed HMM-based map matching algorithm in Section~\ref{sec:semmm}. Section~\ref{sec:eval} presents the experimental evaluation of the \sys{} system. Finally, sections \ref{sec:rw} and \ref{sec:con} discuss related work and conclude the paper,  respectively.

\begin{figure*}[!t]
\centering
\includegraphics[width=0.9\linewidth]{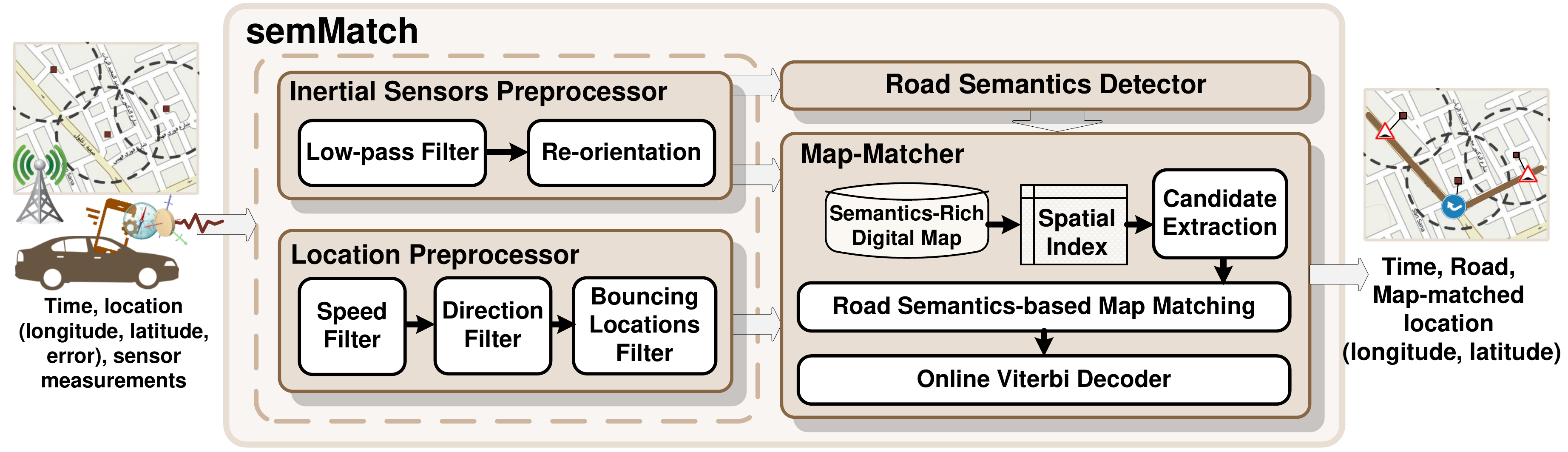}
\caption{\sys{} system architecture.}
\label{fig:arch}
\end{figure*}
\section{semMatch Overview}\label{sec:sys_ov}

In this section, we provide an overview of the \sys{} system and the details of its different supporting modules for handling the noisiness and sparseness of the input location information. We also describe the different road semantics detected by \sys{}. We leave the details of the core of system, \emph{the road semantics-based map matching module}, to the next section.

\subsection{Architecture Overview}

Figure~\ref{fig:arch} shows \sys{} overall architecture diagram. The input to the system is a time-stamped challenging positioning data (e.g. cellular-based localization) along with the phone's available inertial sensors (e.g. accelerometer, magnetometer and gyroscope) measurements. Each input location is represented as a longitude and latitude pair that captures the user's estimated location as well as an estimate of the localization error.

\sys{} starts by preprocessing the input location points to eliminate the noise using the \emph{Inertial Sensor Preprocessor} and the \emph{Location Preprocessor} modules. Then, the \emph{Road Semantics Detector} module identifies the road semantic type when the car passes by them using the inertial sensor measurements.

The preprocessed locations along with the car's current road semantics information are then passed to the \emph{Map Matcher} module that employs a novel incremental HMM-based algorithm that integrates the hints about the car's ambient road semantics and a semantically-enriched digital map to achieve high map matching accuracy in an efficient manner. OpenStreetMaps  have different tags to manually add and query the different road features such as bridges, tunnels, etc. In addition, this semantic tagging can be automated, e.g. using the Map++ system~\cite{aly_map14}.

Our proposed semantic map matching algorithm contains three sub-modules: (1) Candidate Extraction and Filtering, (2) Incremental Road Semantics-based HMM Map Matching, and (3) an Online Viterbi Decoder. The \emph{Candidate Extraction and Filtering} module determines the candidate road segments from the semantically enriched digital map. The module takes into account the error in the input location, the previous estimated user location, and the uncertainty in the semantics detection algorithm. The \emph{Incremental Map Matching} algorithm integrates a number of modifications to the standard HMM map matching algorithm to take the detected semantic type into account as well as the road segment ambient semantics to enhance the accuracy of the estimated road segments and the map matching location on them. Finally, the \textit{Online Viterbi Algorithm} uses dynamic programming to efficiently determine the most probable road segment. The Map Matcher outputs the matched road segment along with the car's estimated location on it (position of the last detected semantic on the road segment).

In the balance of this section, we provide the details of the Pre-processing and the Road Semantics Detector modules. We leave the details of the Map Matching module to the next section.

\begin{figure*}[!t]
  \begin{minipage}[!t]{1.07\textwidth}
\begin{minipage}[!t]{0.35\linewidth}
\centering
  \begin{figure}[H]
\includegraphics[width=\linewidth]{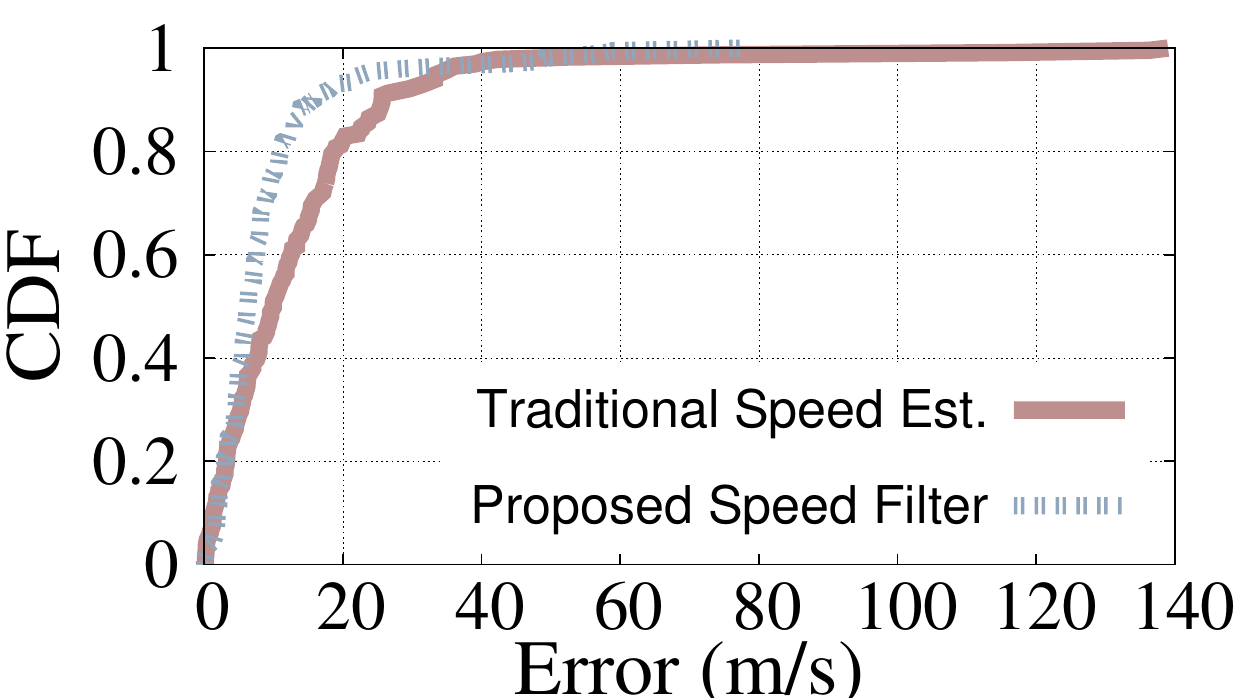}
\caption{Using the new proposed speed filter reduces the median error by more than 36\% as well as reduces the maximum error significantly.}
\label{fig:speedest_ex}
\end{figure}
\end{minipage}
\hspace*{3pt}
\begin{minipage}[!t]{0.6\linewidth}
\centering
\begin{figure}[H]
\centering
    \subfigure[Back-and-forth transitions that can indicate a false turn.]{
      \includegraphics[width=0.35\linewidth]{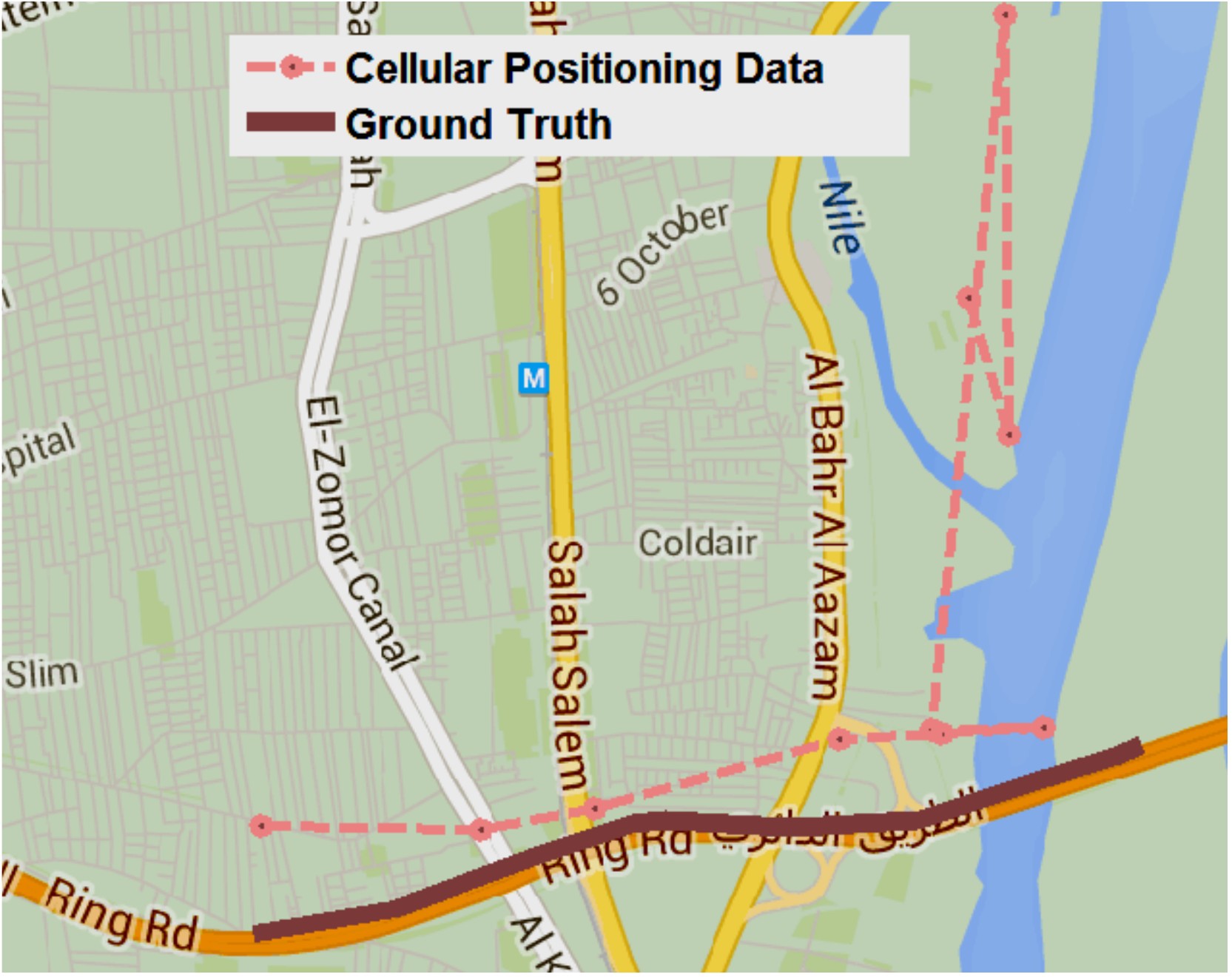}
      \label{fig:map}
    }
\subfigure[The direction change is less than $10^\circ$, indicating that the user is moving in the same direction.]{
      \includegraphics[width=0.55\linewidth]{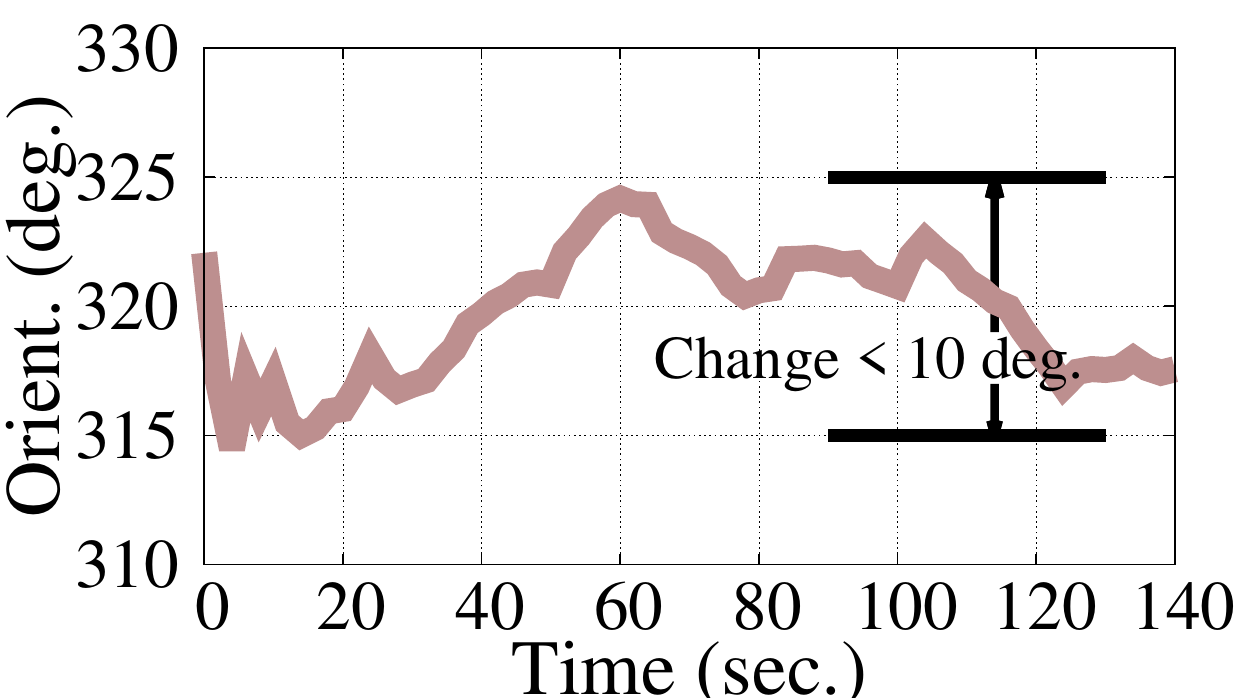}
      \label{fig:sensors}
    }
\caption{Example for back-and-forth transitions in cellular-positioning data indicating a false direction change and how \sys{} detects this outlier using the inertial sensors readings.}
\label{fig:direction_ex}
\end{figure}
\end{minipage}
\vspace{-15pt}
\end{minipage}
  \end{figure*}

\subsection{Preprocessing}
Both the inertial sensors measurements and the input location data are known for their noisy nature~\cite{mohssen2014s}. These are handled by the \textit{Inertial Sensors Preprocessor} and the \textit{Location Preprocessor} respectively.

\subsubsection{Inertial Sensors Preprocessor}
This module is responsible for preprocessing the raw sensor measurements to reduce the effect of (a) phone orientation changes and (b) noise and spurious changes, e.g. sudden breaks, or small changes in the direction while moving.

First, we apply a low-pass filter to the raw sensors data using a weighted local regression filter \cite{cleveland1988locally}. This reduces the noise and spurious changes effect. Then, to handle the orientation changes, we transform the sensors readings from the mobile coordinate system to the world coordinate system leveraging the inertial sensors \cite{mohssen2014s}.
\subsubsection{Location Preprocessor}
For the location data, we apply a series of filters to detect false and misleading transitions and filter them out (Figure~\ref{fig:cell_challenges}). In particular, we apply three filters in succession: the \textit{Speed filter}, the \textit{Bouncing locations filter}, and the \textit{Direction filter}. 

\noindent\textbf{Speed Filter:}

Typically, roads have a maximum speed limit set by traffic regulations. Also, there is a physical maximum speed that cars cannot exceed. Hence, we assume that the car  will not exceed a certain speed threshold ($\nu_{\textrm{max}}$). This threshold is obtained from the digital map and we also add a margin to accommodate drivers that do not strictly follow traffic rules. If a road's speed limit is missing from the map, we set the threshold to the maximum physical speed limits. If the car estimated current speed exceeds this threshold, the new location estimate is detected as an outlier and the car current location estimate remains unchanged.

To estimate the car current speed, intuitively this can be done by dividing the geodesic distance between the car's current location estimate and the new one by the difference in their time-stamps. However, due to the high error in the input locations, this speed is too noisy to use (Figure~\ref{fig:speedest_ex}). Instead, we estimate the car's current speed ($\nu_p$) by averaging the speed between the new location and a window of the preceding preprocessed locations as follows:

\begin{equation}
\nu_p = \frac{1}{w_s}\sum_{i=p-w_s}^{p-1}  d_{i,p}/(t_i-t_p )
\label{eq:speed}
\end{equation}

Where $p$ is the index of the new location estimate, $w_s$ is the window size, $d_{i,p}$ is the geodesic distance between locations $i$ and $p$, and $t_i, t_p$ are the time-stamps of locations $i$ and $p$ respectively.
\\

\noindent\textbf{Bouncing Locations Filter:} 
To reduce the back-and-forth transitions, we apply an $\alpha$-trimmed mean filter~\cite{mohamed2014accurate} on the location points. An $\alpha$-trimmed filter has the advantage of handling both impulse and Gaussian noise, as compared to mean and median filters that can handle only one of them. Moreover, it is simple to implement.

The basic idea  is to look at the neighbors of each point, remove $2\alpha$ of the extreme neighbors, i.e. outliers, then replace the point by calculating the mean of the unfiltered neighbours.  Therefore, at $\alpha=0$, the filter works as a standard average filter while at $\alpha=0.5$, the filter works as a median filter.

Note that to apply the filter, we need to sort the locations ($\textrm{loc}_i$). We experimented with different space-filling curves and found that applying the linear space filling curve \cite{bader13space-filling} provides good accuracy while maintaining low computational time. This is intuitive as most of the time the car moves in straight lines.
\\\\
\noindent\textbf{Direction Filter:}\\
Back-and-forth transitions can mislead the map matching algorithm to sense that the user made a change in her direction, i.e. as if the user took a u-turn. Therefore, to reduce this effect we apply the direction filter. It ensures that the change in the car direction is only allowed when we are sure that it is originating from an actual change in direction, not due to the input data noisiness.

To do that, we leverage the smartphone's inertial sensors to estimate the car heading direction~\cite{mohssen2014s}. Specifically, if a location point indicates a direction change, we confirm this change using the car estimated bearing from the inertial sensors. If the indicated direction change proved false, the car raw location remains unchanged. Figure~\ref{fig:direction_ex} shows an example where the input cellular-location data indicates a direction change, while the inertial sensors shows that it is a false one.

\begin{figure*}[!t]
  \begin{minipage}[!t]{1.07\textwidth}
\begin{minipage}[!t]{0.6\linewidth}
\centering
   \begin{figure}[H]
\centering
    \subfigure[Going over a speed bump leads to a large change in the phone's gravity acceleration.]{
      \includegraphics[width=0.45\linewidth]{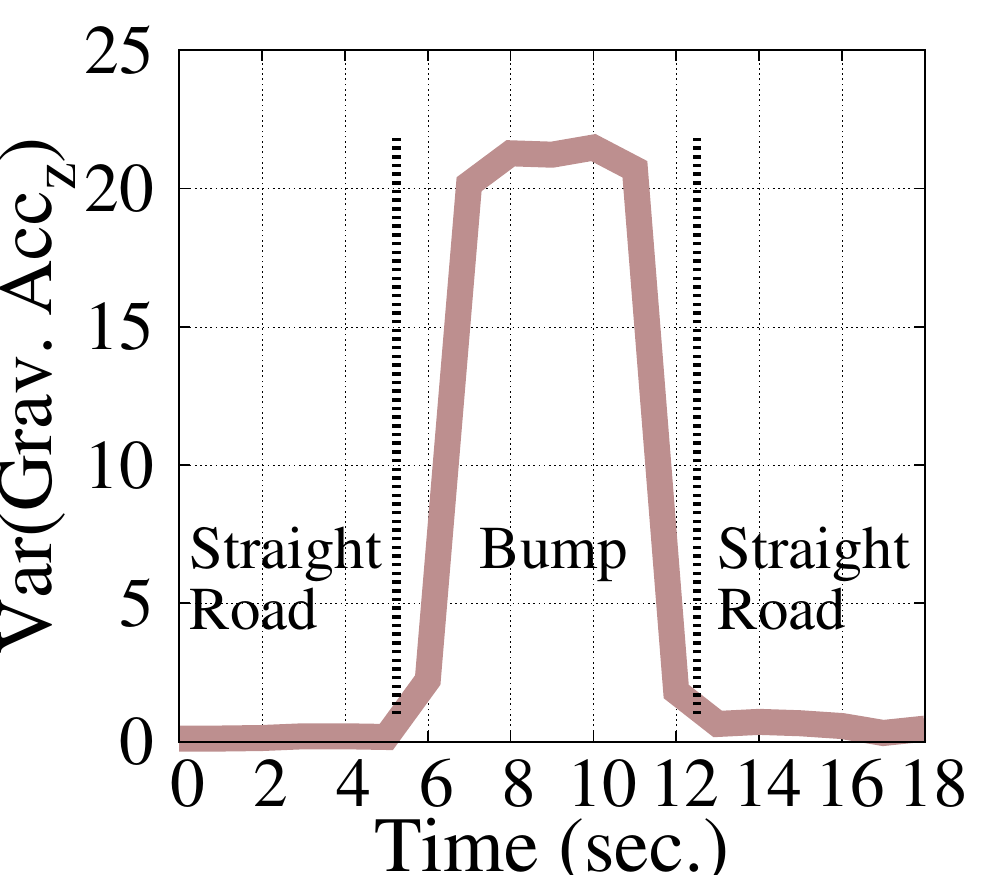}
      \label{fig:bump_ex}
    }
\subfigure[Taking a u-turn leads to a change in the user direction (i.e. phone orientation) around 180$^\circ$.]{
      \includegraphics[width=0.45\linewidth]{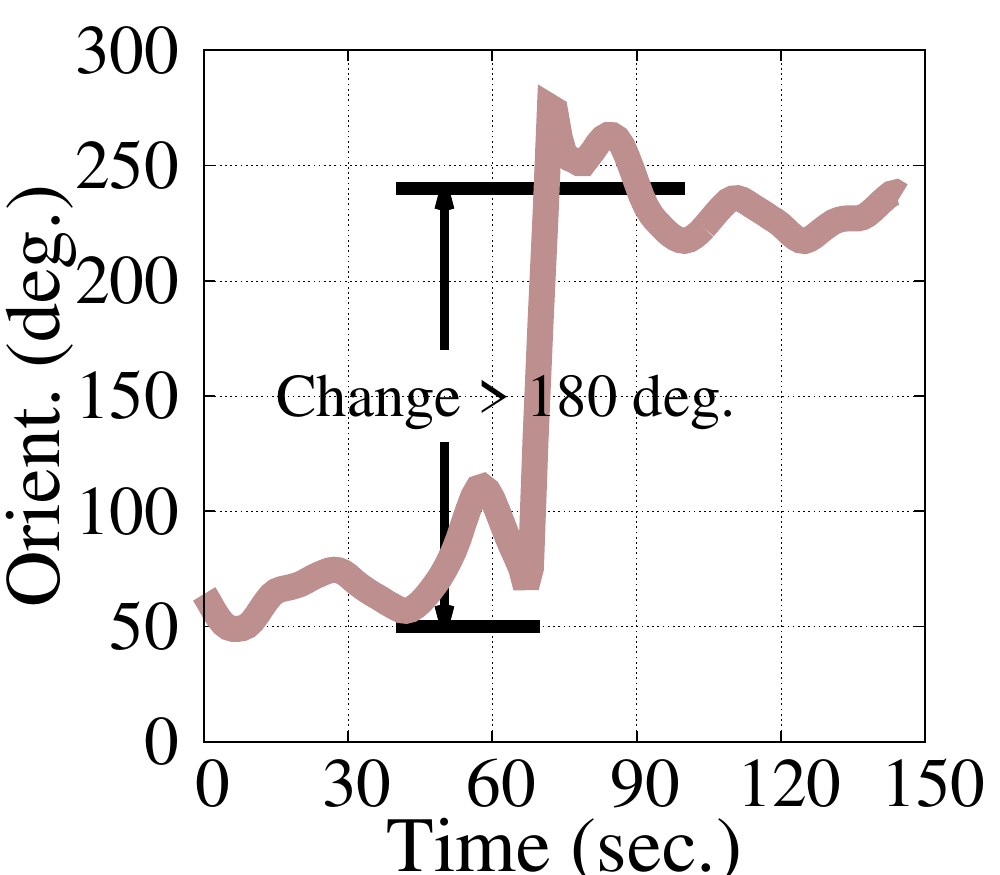}
      \label{fig:uturn_ex}
    }
\caption{Passing by the different road semantics (e.g. speed bumps and u-turns) has unique effects on the smartphone's sensors measurements.}
\label{fig:sem_example}
\end{figure}

\end{minipage}
\hspace*{3pt}
\begin{minipage}[!t]{0.32\linewidth}
\centering
\begin{figure}[H]
\centering
\includegraphics[width=\linewidth]{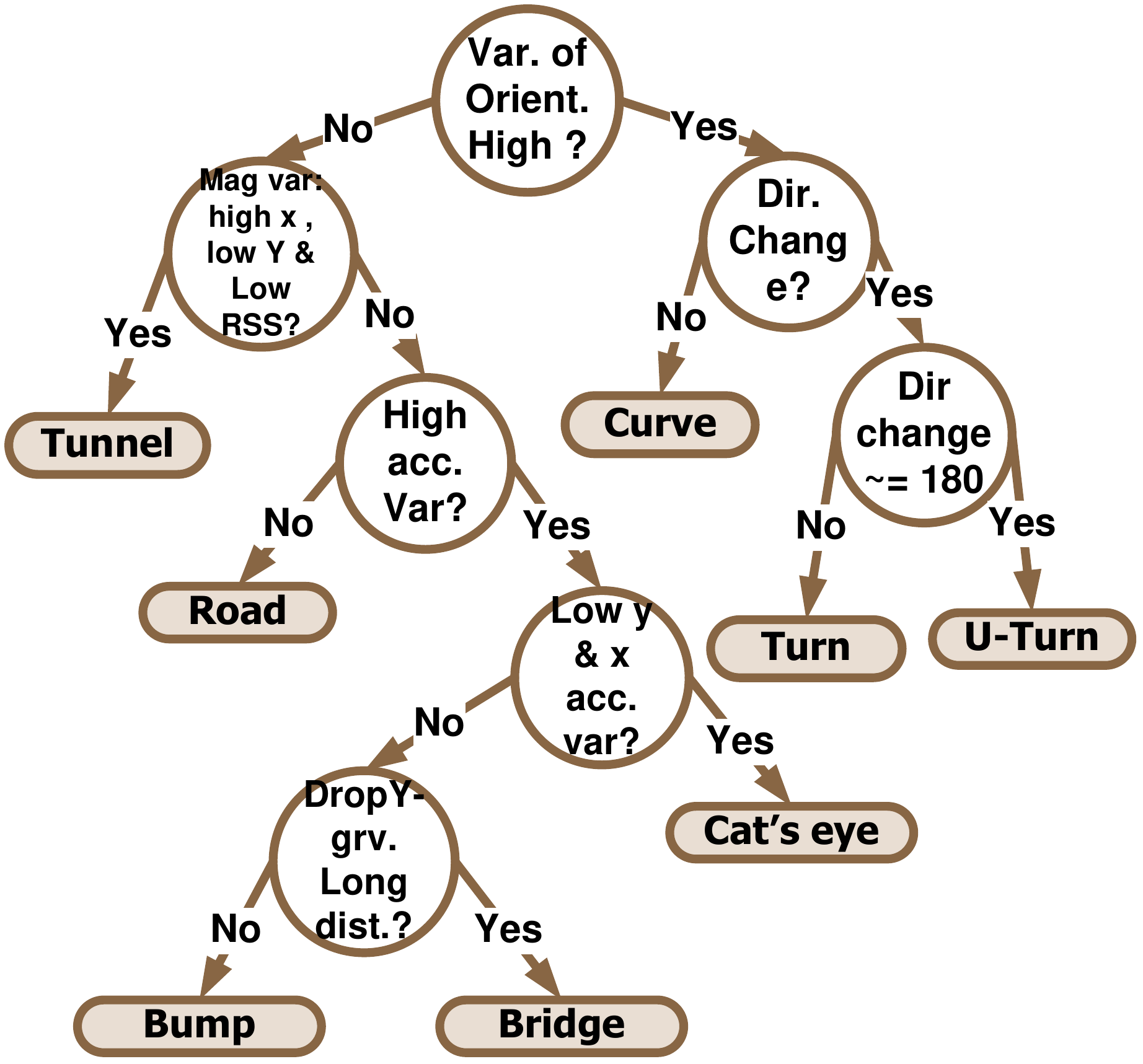}
\caption{Decision tree used by \sys{} to identify the different road semantics in real-time.}
\label{fig:sem_det}
\end{figure}
\end{minipage}
\vspace{-15pt}
\end{minipage}
  \end{figure*}

\subsection{Road Semantics Detector}\label{sec:tree}

\begin{table}[!t]
\caption{Confusion matrix for classifying different map semantics discovered from {\bfseries in-vehicle} traces.}
\centering
\small{
\begin{tabular}{|c||c c c c c c c c||} \cline{2-9}
\multicolumn{1}{c||}{}	& \rot{Cat's eye}& \rot{Bump}	&\rot{Curve}&\rot{Bridge}	&\rot{Tunnel}&\rot{Turn} &\rot{U-turn} &\ret{No Class}\\\hline\hline
Cat's eye	& \cellcolor{gray!10}{\bfseries 22}&0&0&0&0&0&0&5\\\hline
Bump	& 0&\cellcolor{gray!10}{\bfseries 37}&0&0&	0&0&0&0\\\hline
Curve	&0&0&\cellcolor{gray!10}{\bfseries 33}	&0&0&0&0&0\\\hline
Bridge	&0&0&0&\cellcolor{gray!10}{\bfseries 11	}&0&0&0&3\\\hline
Tunnel	&0&0&0&0&\cellcolor{gray!10}{\bfseries 15}&0&0&0\\\hline
Turn	&0&0&0&0&0&\cellcolor{gray!10}{\bfseries 55}&0&0\\\hline
U-turn	&0&0&0&0&0&0&\cellcolor{gray!10}{\bfseries 12}&0\\\hline\hline

\end{tabular}
}
\label{tab:drive_conf}
\end{table}

The \emph{Road Semantics Detector} module processes the smartphone's inertial sensor measurements to detect when the car passes by a road semantic and its type. In~\cite{aly_map14}, we found that the smartphone sensors get affected by the different road features such as speed bumps or tunnels. For example, a car going over a speed bump will experience an up-and-down movement over a small distance which leads to a high variance in the phone's measured gravity acceleration (Figure~\ref{fig:bump_ex}). Similarly, a car taking a u-turn will experience a large change in its direction (around 180$^\circ$) as shown in Figure~\ref{fig:uturn_ex}. Hence, we use these effects to detect them.

\sys{} identifies a wide range of road semantics in real-time using a decision-tree classifier (Figure~\ref{fig:sem_det}) including \textbf{tunnels, curves, bridges, turns, speed bumps, u-turns, and cat's eyes}. We refer the readers to~\cite{aly_map14} for more information about the features we used to identify the different road semantics using the smartphone's sensors. Table~\ref{tab:drive_conf} shows the confusion matrix between the different classes. The different road semantics can be identified with high accuracy due to their unique signatures. However, to compensate for cases not observed during the classifier training, we add a small error to every cell in the matrix to account for further confusion.

\section{Semantic Map Matching}\label{sec:semmm}
Through this section, we provide the details of our novel road semantics-based \emph{Map Matcher} module. We first start by providing an overview on the new model and its differences from previous HMM algorithms. Then, we provide the mathematical model and the details of the HMM model components.
\subsection{Overview}
In traditional map matching using GPS traces, e.g. \cite{newson2009hidden}, the input to the system is a trace of the noisy GPS locations and the output is the corresponding sequence of road segments the car passed by. Extending this directly to the coarse-grained, noisy, and sparse cellular location information (Figure~\ref{fig:cell_challenges}) leads to poor map matching accuracy with high complexity as explained in Section~\ref{sec:intro}. \sys{} solves these challenges by leveraging the road semantics the car passes by to reduce ambiguity and hence enhances both the map matching accuracy and running time. Specifically, the input to the map matcher module is a sequence of ordered triples in the form of \textbf{(Coarse-grained location, Estimated location error, Semantic type)} representing the road semantics detected by the phone sensors during the car movement at a specific coarse-grained input location. Note that due to the rich number of semantics detected by \sys{} (as described in Section~\ref{sec:tree}), the input data to \sys{} is much more frequent than, e.g. a sequence of unique cell IDs only, and hence should lead to better accuracy and fewer road segments candidates as described in details in the rest of this section.
\begin{figure*}[!t]
  \begin{minipage}[!t]{1.07\textwidth}
\begin{minipage}[!t]{0.25\linewidth}
\centering
\begin{figure}[H]
\centering
\includegraphics[width=\linewidth]{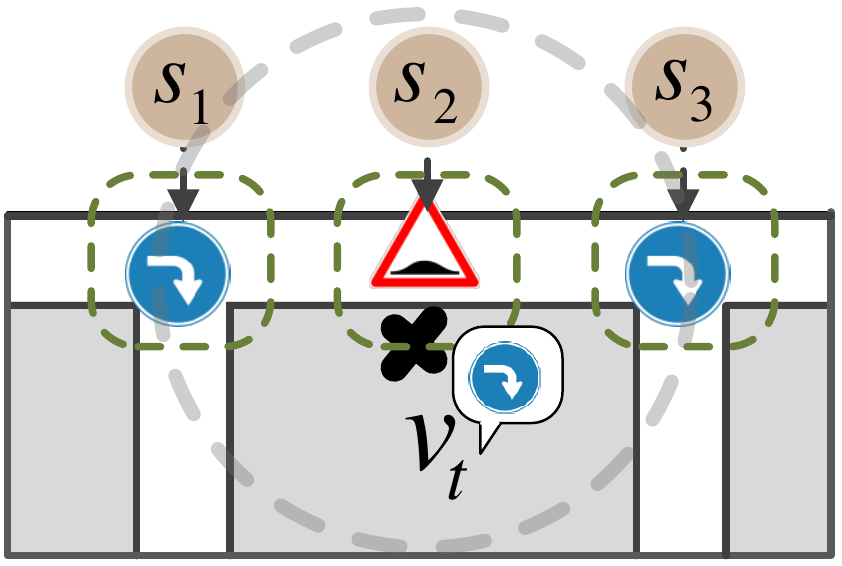}
\caption{Proposed road semantics-based HMM model. A state ($s_i$) represents a road segment semantic landmark and is modelled as an ordered triples <Physical road segment ID, Landmark type (e.g. bump in state $s_2$), Landmark location on the road>. The user observations ($v_t$'s) are defined by the estimated location (x mark), the detected semantic type (e.g. turn in $v_t$), and the estimated error (dotted circle). The observation probability ($p(v_t|s_i)$) is a function of both the road landmark's location and type and the observation's location and type.}
\label{fig:hmm_states}
\end{figure}
\end{minipage}
\hspace*{3pt}
\begin{minipage}[!t]{0.32\linewidth}
\centering
\begin{figure}[H]
\center
\includegraphics[width=0.8\linewidth]{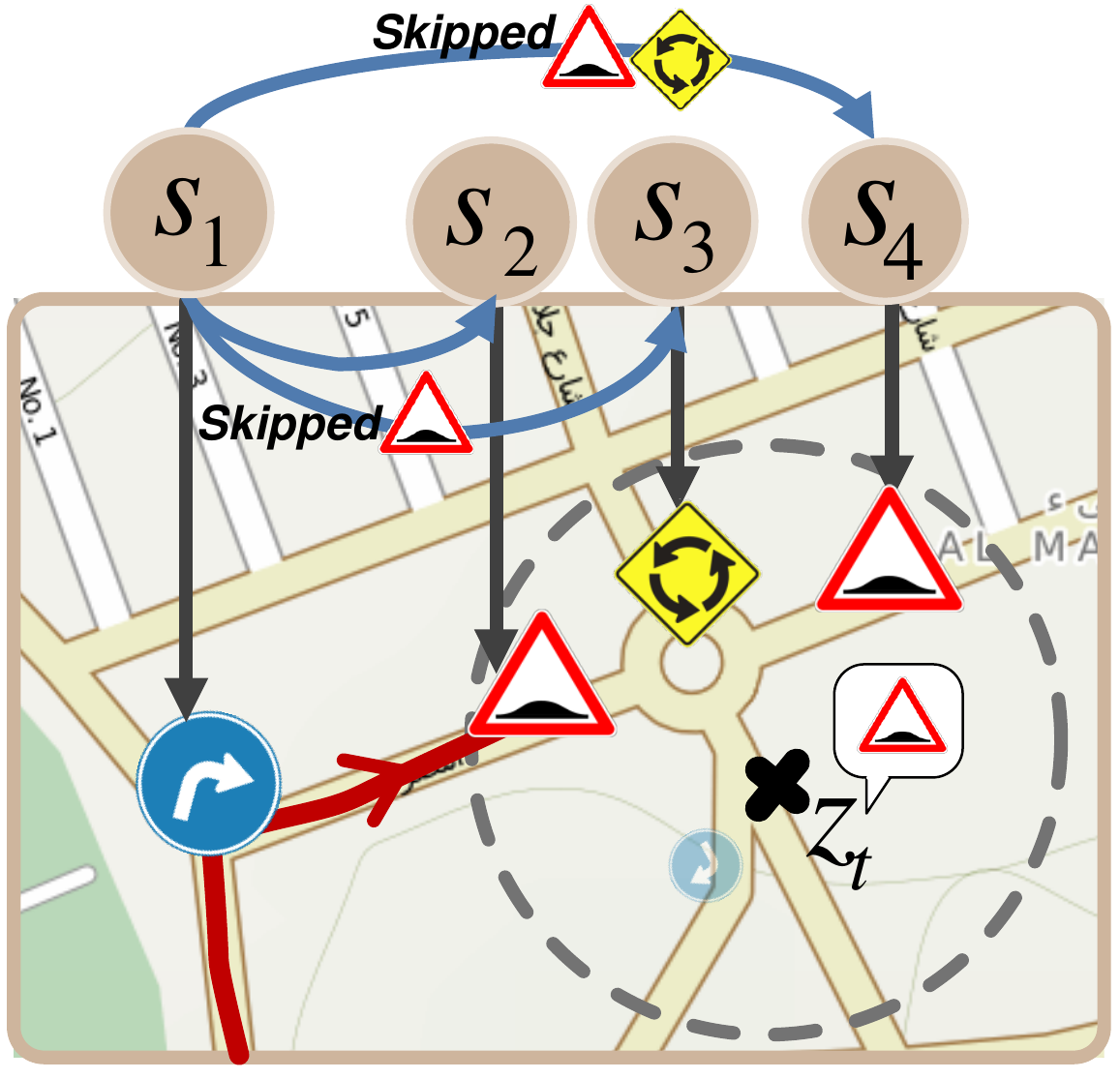}
\caption{Transition probabilities calculation from state $s_1$ to states $s_2$, $s_3$, and $s_4$. The transition probability is a function of 1) the difference in the heading between the hidden states' road segments on the map and the car change of heading as sensed by the phone sensors and 2) the skipped landmarks on the map. States $s_3$ and $s_4$ are penalized due to skipping one or more road semantics. } \label{fig:trns_ex}
\end{figure}

\end{minipage}
\hspace*{3pt}
\begin{minipage}[!t]{0.35\linewidth}
\centering
\begin{figure}[H]
\centering
      \includegraphics[width=\linewidth]{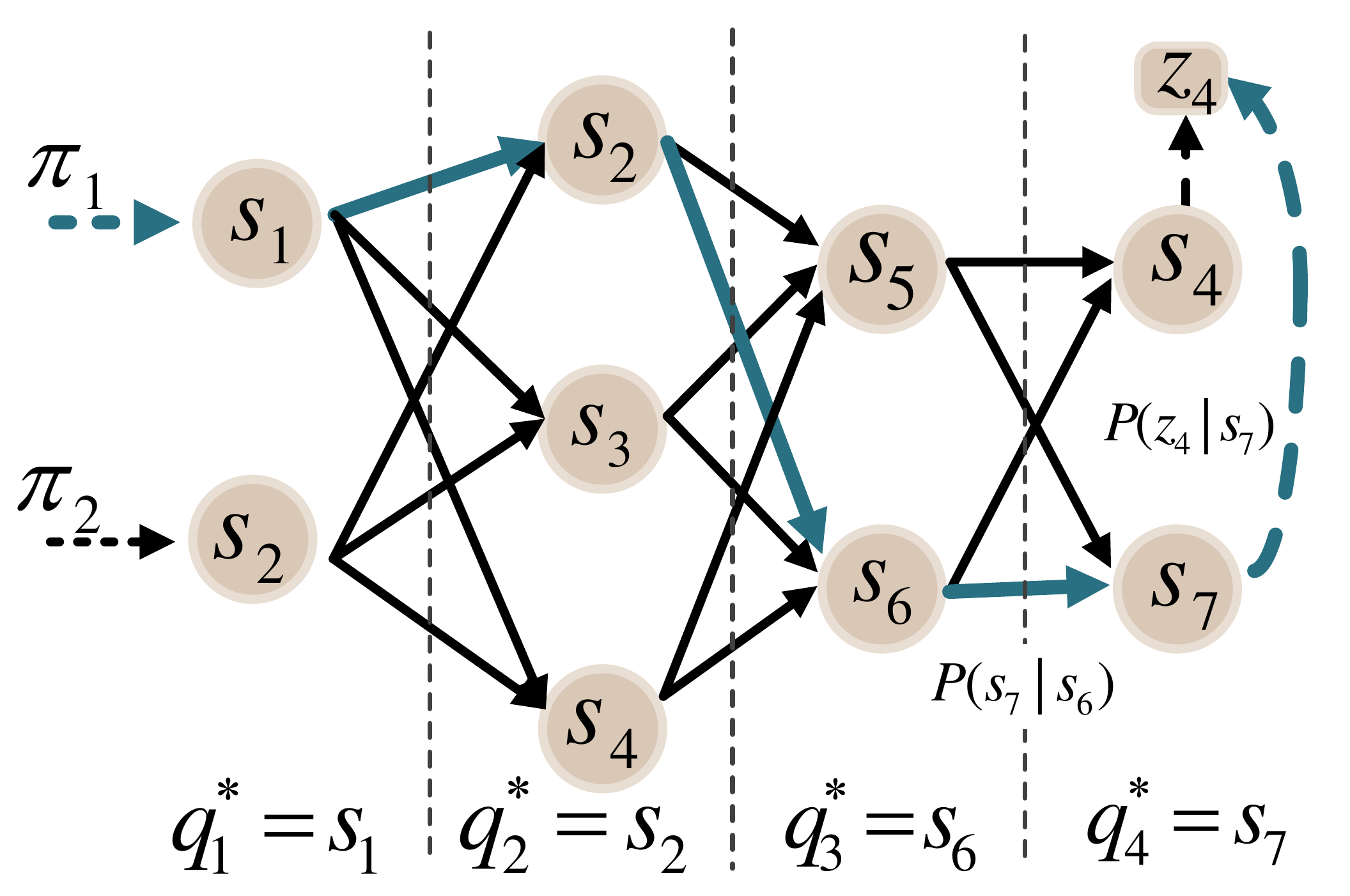}
\caption{Illustration of the proposed HMM model Viterbi decoding: States are the graph vertices, states transition and observation probabilities are drawn as links in the graph. The initial states probabilities ($\pi_i$'s) are estimated for the first set of states in the current sequence. The Viterbi decoded sequence ($Q$) is colored in blue. The user observation is drawn here only at time $t=4$ due to the incremental nature of the algorithm.}
\label{fig:viterbi}
\end{figure}
\end{minipage}
\vspace{-15pt}
\end{minipage}
  \end{figure*}
\subsection{Mathematical Model}

A HMM can be represented as $\lambda= (S, V, A, B, \pi)$
\cite{hmm} where:

\begin{itemize}
\item $S=\{s_1,...,s_{N}\}$ is the set of possible states and $N= |S|$.
In our case, each state represents a road semantic landmark (e.g. a speed bump) on a specific road segment as shown in Figure~\ref{fig:hmm_states}. Therefore, a state $s_i$ is represented by the ordered triples \textbf{(Road segment ID (ID), Road semantic landmark type (type), Road semantic landmark location (loc))}, where the ``Road semantic landmark location'' represents the coordinates of the semantic landmark on the given road segment. \textbf{\emph{Note that using this representation, each physical road segment is split into a number of states by the different semantic landmarks on this segment}} (Figure~\ref{fig:hmm_states}).

\item $V=\{v_1,...,v_M\}$ is the set of observations from the
model  and $M= |V|$. In our case, each observation is a detected road semantic type using the phone inertial sensors and is represented by an ordered quadruple in the form of \textbf{(Coarse-grained location (loc), Estimated location error ($\textrm{err}$), Estimated heading direction ($\theta$), Detected semantic type (type))}.

\item $A= \{a_j(k)\}$ is the observation symbol probability
distribution in state $j$, where $a_j(k)= P[z_t=v_k\, |q_{t}= s_j], i<j< N, 1<k<M$ and $z_t$ and $q_t$ are the observation and state at time $t$, respectively. In other words, this represents the probability of observing a certain road semantic type given the car is at a specific state, i.e. road segment with a certain road semantic landmark.

\item $B= \{b_{ij}\}$ is the state transition probability
distribution, where $b_{ij}= P[q_{t+1}= s_j|q_{t}= s_i], i, j< N$.

\item $\pi=\{\pi_i\}$ is the initial state distribution, where $\pi_i=P[q_1=s_i]$.
\end{itemize}

 Therefore, the problem becomes, given a sequence of observations $Z = (z_1,...,z_T)$, where $T$ is a system parameter, and each $z_i \in V, 1<i<T$,
  we want to find the most probable sequence of semantic road segments (states) $Q
=(q^*_1,...,q^*_T)$, where each $q^*_i \in S, 1<i<T$. In the rest of this section, we give the details of how \sys{} models these different  components.
\subsection{Observation Probability Distribution (A)}
Since there can be ambiguity in the detected semantics due to the noisy phone sensors, e.g. a cat's eye detected as a speed bump, the probability of the phone detecting a given road semantic $Y$ (e.g. a speed bump) given a specific road segment semantic landmark $X$ (e.g. a cat's eye) should be a function of two factors (Figure~\ref{fig:hmm_states}): 1) the distance between the location of the car where the semantic $Y$ was detected and the location of the semantic landmark $X$ on the road segment. The higher this distance, the lower the probability that should be assigned to a given road segment. 2) the probability of the phone sensors detecting type $Y$ when the car is passing by a semantic of type $X$. This is captured by the false positive and false negative rates of detecting the different semantics as in Section~\ref{sec:tree}.

\sys{} models the \emph{first factor} as a Gaussian distribution:
\begin{equation}
f_1 = \frac{1}{\sqrt{2\pi}z.\sigma}e^{-0.5\left(\frac{\textrm{dist}(z.loc,s_{i}.loc)}{z.\sigma}\right)^2}
\end{equation}
Where $\textrm{dist}(z.loc,s_{i}.loc)$ is the geodesic distance between the observation $z$ and the road segment semantic, i.e. hidden state,  $s_{i}$. $z.\sigma$ is the standard deviation of a Gaussian random variable that corresponds to the error in the observed input location.

For the \emph{second factor}, we model it using the \textit{Road Semantics Detector} confusion matrix to estimate the probability of observing a semantic type $z.type=Y$ at time $t$ conditioned on the hidden state $s_{i}$ road semantic landmark type ($s_i.type=X$):
\begin{equation}
f_2 = p(z.type|s_i.type)
\label{eq:weight_model}
\end{equation}

Hence, the final observation probability, $p(z|s_{i})$, is modeled as:
\begin{equation}
p(z|s_{i}) = p(z.\textit{type}|s_i.\textit{type})\frac{1}{\sqrt{2\pi}z.\sigma}e^{-0.5\left(\frac{\textrm{dist}(z.loc,s_{i}.loc)}{z.\sigma}\right)^2}
\label{eq:emission_pr}
\end{equation}

\subsection{Transition Probability Distribution (B)}
The transition probability is the probability of moving to the next state $s_{j,t}$ given the current state is  $s_{i,t-1}$.
Intuitively, for a probable transition between two road segment landmarks, the change in the car orientation (as detected by the phone sensors) between the two segments should match the change of the orientation in the digital map for the same two segments.

Therefore, \sys{} models this intuition by a function of the heading change in the input observations ($\Delta\theta_z$) and the heading change between the hidden states pair ($\Delta\theta_s$) using a Gaussian distribution as:
\begin{equation}
f(\Delta\theta_z,\Delta\theta_s) = \frac{1}{\sqrt{2\pi}\sigma_h}e^{-0.5\left(\frac{|\Delta\theta_z - \Delta\theta_s|}{\sigma_h}\right)^2}
\end{equation}
Where $\sigma_h$ is the estimated standard deviation of the heading change error. The more similar the change in heading direction, the higher the transition probability. However, in some cases, due to the high error in the input location, we can have more than one hidden state with a similar semantic type (Figure~\ref{fig:trns_ex}). However, misleading semantics will lead to skipping road segment landmarks. Therefore, to capture this in the transition probability, we penalize transitions that skip one or more road semantics. The final transition probability in \sys{} is estimated as:
\begin{equation}
p(s_{j}|s_{i}) = \frac{1}{\sqrt{2\pi}\sigma_h}e^{-0.5\left(\frac{|\Delta\theta_z - \Delta\theta_s|}{\sigma_h}\right)^2} \prod_{l \in L} p(\bar{l}|l)
\end{equation}
Where $L$ is the set of skipped road semantic segments between the states pair $s_{j}$ and $s_{i}$. For a semantic $l \in L$, the probability $p(\bar{l}|l)$ represents the probability of passing by the semantic $l$ and not detecting it. We model this probability using the false negative value of the semantic $l$ computed from the confusion matrix of our \emph{Road Semantics Detector} module (Table~\ref{tab:drive_conf}).

To estimate $\sigma_h$, we use the ground-truth data. Note that, unlike the input positioning data which has highly dynamic and variable accuracy, the sensors heading accuracy has a more consistent accuracy outdoors~\cite{aly2015lanequest}. Hence, we use a fixed $\sigma_h$ calculated based on the Median Absolute Deviation (MAD) of the direction changes in our ground truth data, which is a robust estimator of  $\sigma_h$~\cite{aly2015lanequest,newson2009hidden}:
\begin{equation}
\sigma_h = 1.4826 \times median(|\Delta\theta_{z} - \Delta\theta_{s}|)
\end{equation}

\subsection{Initial State Distribution}
At each time instance, there is a large number of candidate road semantic segments that the map matcher can work with, which affects both the system accuracy and running time negatively. To reduce the search space, the \emph{Candidate Extraction and Filtering} module leverages the the current coarse-grained cellular input location estimate ($\textit{loc}$), the associated estimated error in this location ($\textit{err}$), and the last estimated candidate road segments. In particular, the module works in two steps: First, it extracts the candidate road semantics from the digital map that fall inside the circle centered at $\textit{loc}$ with radius  $\textit{err}$. To speed up this process, it builds an R-tree spatial index~\cite{guttman1984r} on all possible road semantic segments in the road network.
Second, it remove candidate road semantic segments that are not connected to any of the candidate road segments from the previous estimation step.

Initially, all road semantics landmarks selected by the \emph{Candidate Extraction and Filtering} module are assigned weights based on their semantic type as:
\begin{equation}
  \pi_{i,1} = p(z_1.\textit{type}|s_i.\textit{type})
  \label{eq:prior0}
\end{equation}
which is obtained from the semantic types confusion matrix. This is intuitive as if the user observed a bump, the states with a bump as their landmark should have initial probability higher than states with other semantic types (e.g. tunnels).

The initial state distribution is re-estimated after each step in our semantics HMM for the new states as the product of the old initial state distribution and the state's transition probability as:
\begin{equation}
\pi_{i,t} =  \sum_{j \in S_{t-1}} \pi_{j,t-1}p(s_{i,t}|s_{j,t-1})
  \label{eq:prior1}
\end{equation}

\subsection{Optimal State Sequence Estimation}
The Viterbi decoder algorithm aims to determine the most probable hidden states sequence based on the estimated HMM probabilities. The selected hidden states represent the car's traversed road semantic segments.

For \sys{} to operate in real-time, it cannot wait till the whole sequence is available. Hence, in an incremental manner, \sys{} uses a sliding window on the \emph{Map Matcher} module input locations. Every time a new location is introduced, the HMM parameters are calculated for the new introduced location and the associated candidate states. 
The online Viterbi algorithm \cite{bloit2008short,vsramek2007line} is then applied to compute the maximum likelihood sequence of hidden states for the current window using dynamic programming by extending the current solution with the estimated HMM parameters to get the best path (Figure~\ref{fig:viterbi}). Also, the initial states probability is updated as the sliding window moves.

Algorithm~\ref{alg:mapmatch} summarizes the \sys{} map matching algorithm.

\begin{algorithm}[!t]
\caption{\sys{} Map Matching Algorithm}
\begin{algorithmic}
\State \textbf{Input:} $z_t$\Comment{User observed point at time $t$}
\State \textbf{Output:} $loc,id$\Comment{map matched user location and road segment id}
\State Let $S = \{s_1,s_2,...,s_N\}$ be the hidden states set.
\State Let $S_{t-1}[]$ be the previous hidden states vector set;
\State Let $\Pi_{t-1}[]$ be vector set for the states' initial probability at time $t-1$
\State \textbf{Procedure:}
\State $S_t$ $\gets$ \textbf{\textit{ExtractCandidates($S,z_t$)}}\\\Comment{Extract candidates road semantics}
\For{\textbf{each} $s_i \in S_t$}
\State ob[i] $\gets$ $p(z_t.\textit{type}|s_i.\textit{type})\frac{1}{\sqrt{2\pi}z_t.\sigma}e^{-0.5\left(\frac{\textrm{dist}(z_t.\textit{loc},s_i.\textit{loc})}{z_t.\sigma}\right)^2}$
\\\Comment{Calculate Observation Probability}
\If{\textbf{\textit{isEmpty($S_{t-1}$)}}}
\State $\Pi_{t-1}[i]$ $\gets$ $p(z_t.\textit{type}|s_i.\textit{type})$
\Else
\For{\textbf{each} $s_j \in S_{t-1}$}
\State $L$ $\gets$ \textbf{\textit{SemanticsSearch($S,s_j,s_i$)}}
\\\Comment{Search for semantics between $s_i$ and $s_j$}
\State FN $\gets$ 1
\For{\textbf{each} $l \in L$}
\State FN $\gets$ FN * $p(\bar{l}|l)$
\EndFor
\State tr[i][j] $\gets$ $\frac{1}{\sqrt{2\pi}\sigma_h}e^{-0.5\left(\frac{|\Delta\theta_z - \Delta\theta_s|}{\sigma_h}\right)^2} \times$ FN
\\\Comment{Calculate Transition Probability}
\EndFor
\EndIf
\EndFor
\State Q,$\Pi_{t-1}$ $\gets$ \textbf{\textit{OViterbiDecoder($\Pi_{t-1}$[],tr[][],ob[])}}\Comment{Apply Online Videtrbi Decoder to get the most probable sequence and the updated initial probability}
\State \textbf{\textit{output}} $q_t.\textit{loc},q_t.\textit{ID}$ \\ \Comment{Location  and road id of last state in the sequence Q}
\end{algorithmic}
\label{alg:mapmatch}
\end{algorithm}

\begin{figure*}[!t]
  \begin{minipage}[!t]{1.07\textwidth}
\begin{minipage}[!t]{0.3\linewidth}
\centering
    \begin{figure}[H]
\centering
      \includegraphics[width=\linewidth]{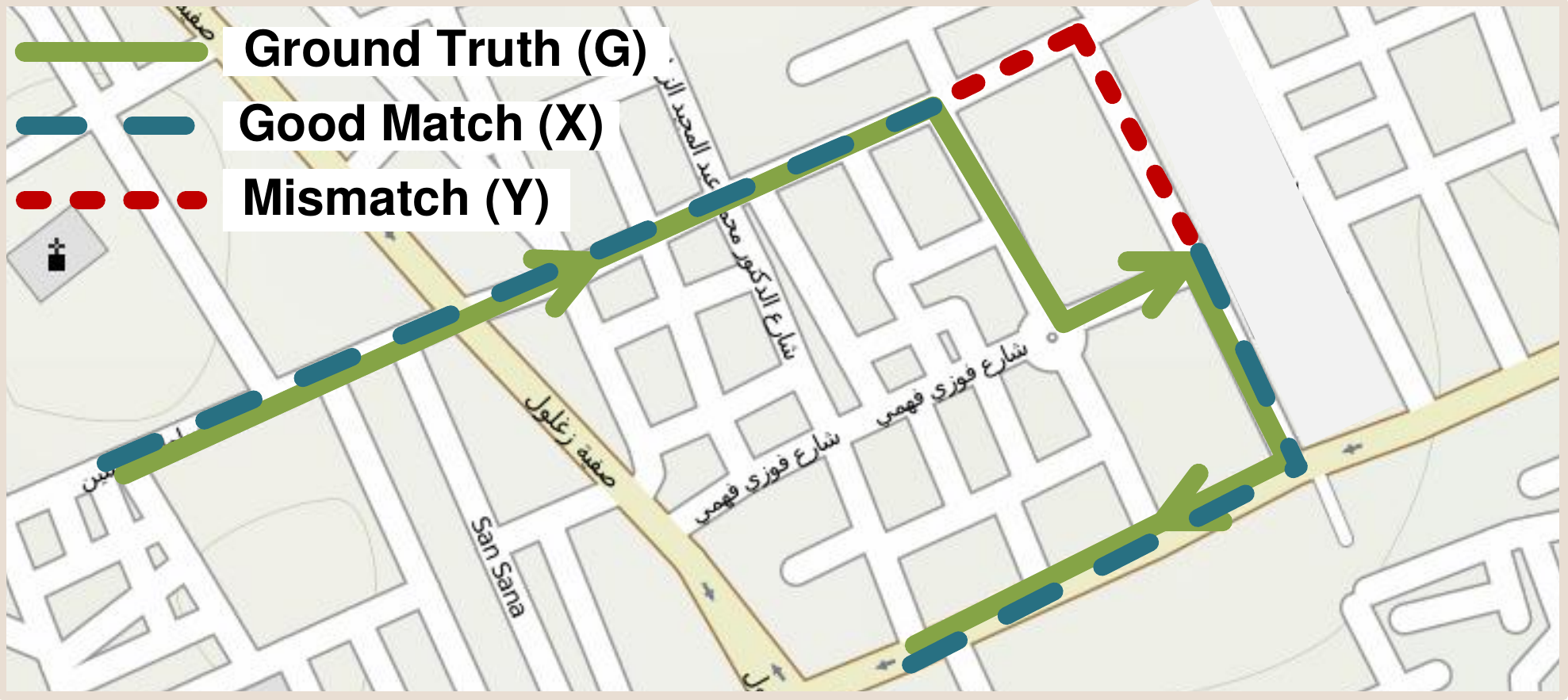}
\caption{An example showing the definition of precision ($X/(X+Y)$) and recall ($X/G$).}
\label{fig:metrics}
\end{figure}
\end{minipage}
\hspace*{3pt}
\begin{minipage}[!t]{0.3\linewidth}
\centering
\begin{figure}[H]
\centering
      \includegraphics[width=\linewidth]{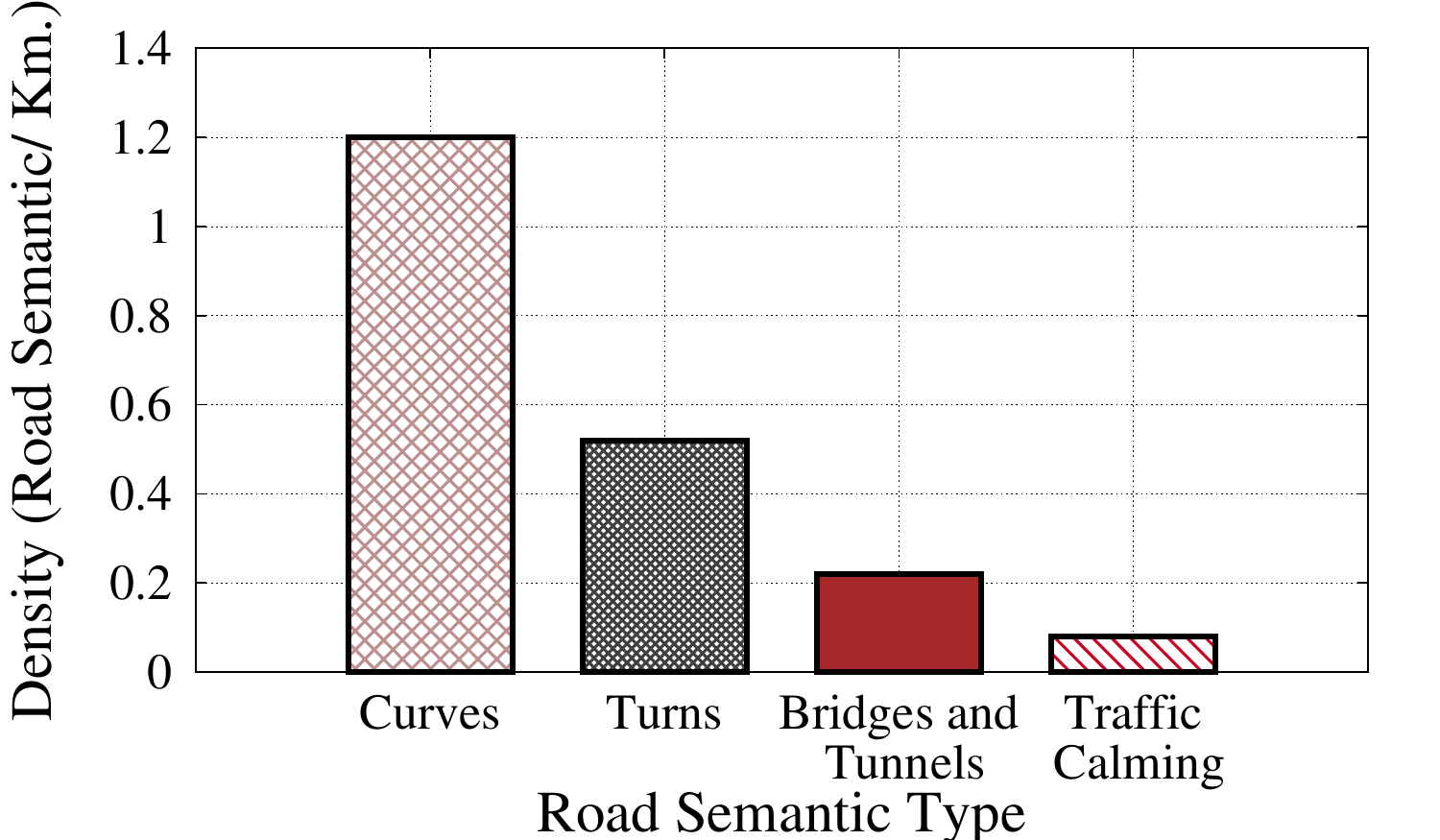}
\caption{Density of the different road semantics in our trajectory traces.}
\label{fig:semdensity}
\end{figure}
\end{minipage}
\hspace*{3pt}
\begin{minipage}[!t]{0.3\linewidth}
\centering
\begin{figure}[H]
\centering
      \includegraphics[width=\linewidth]{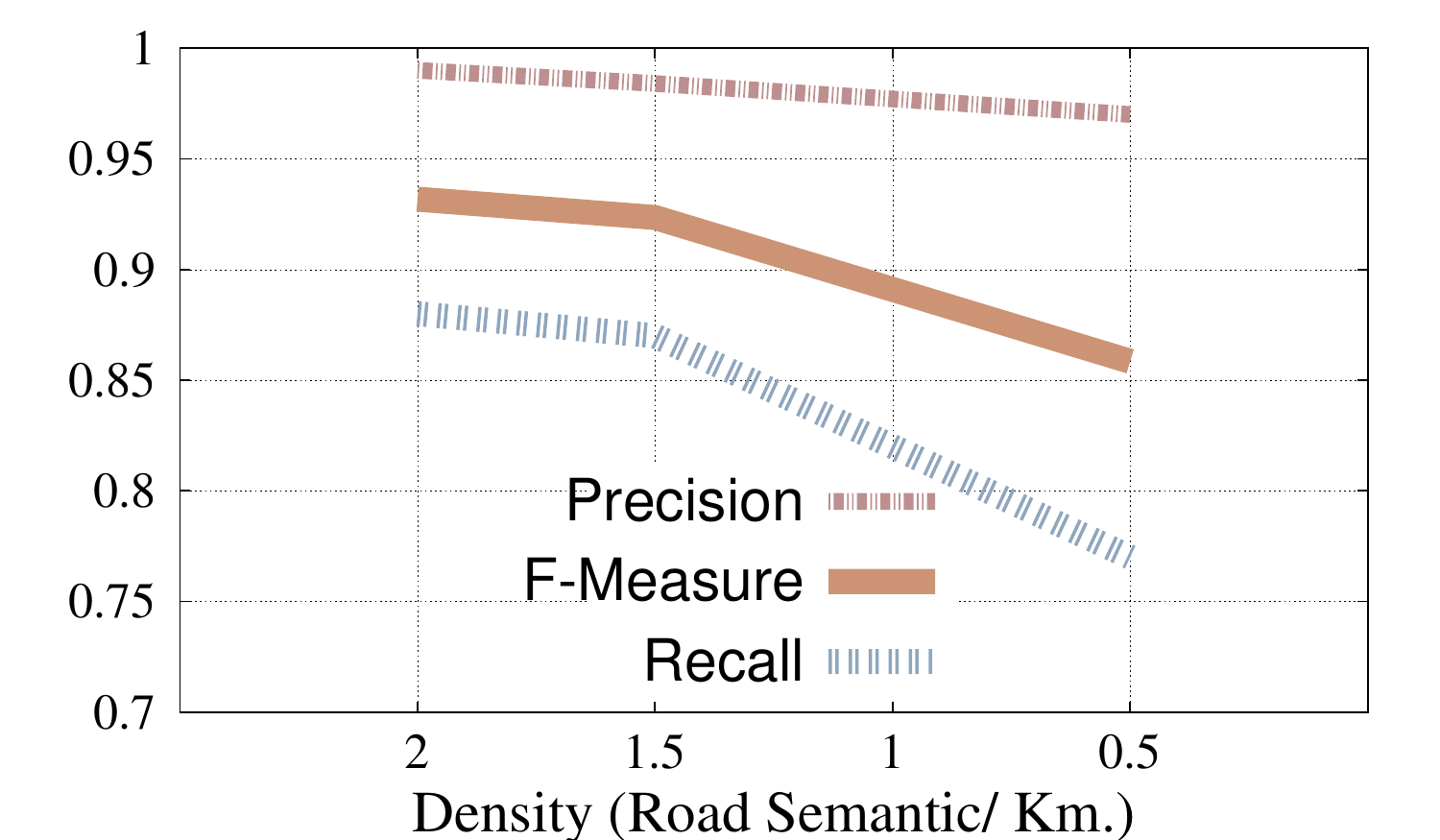}
\caption{Performance of \sys{} under different road semantics density.}
\label{fig:density}
\end{figure}

\end{minipage}
\vspace{-15pt}
\end{minipage}
  \end{figure*}
\begin{figure*}[!t]
\centering
 \subfigure[Cellular Pos. Data (average pos. accuracy=1.9km,
avg. update rate= 1.4 loc. estimate per km)]{
      \includegraphics[width=0.3\linewidth]{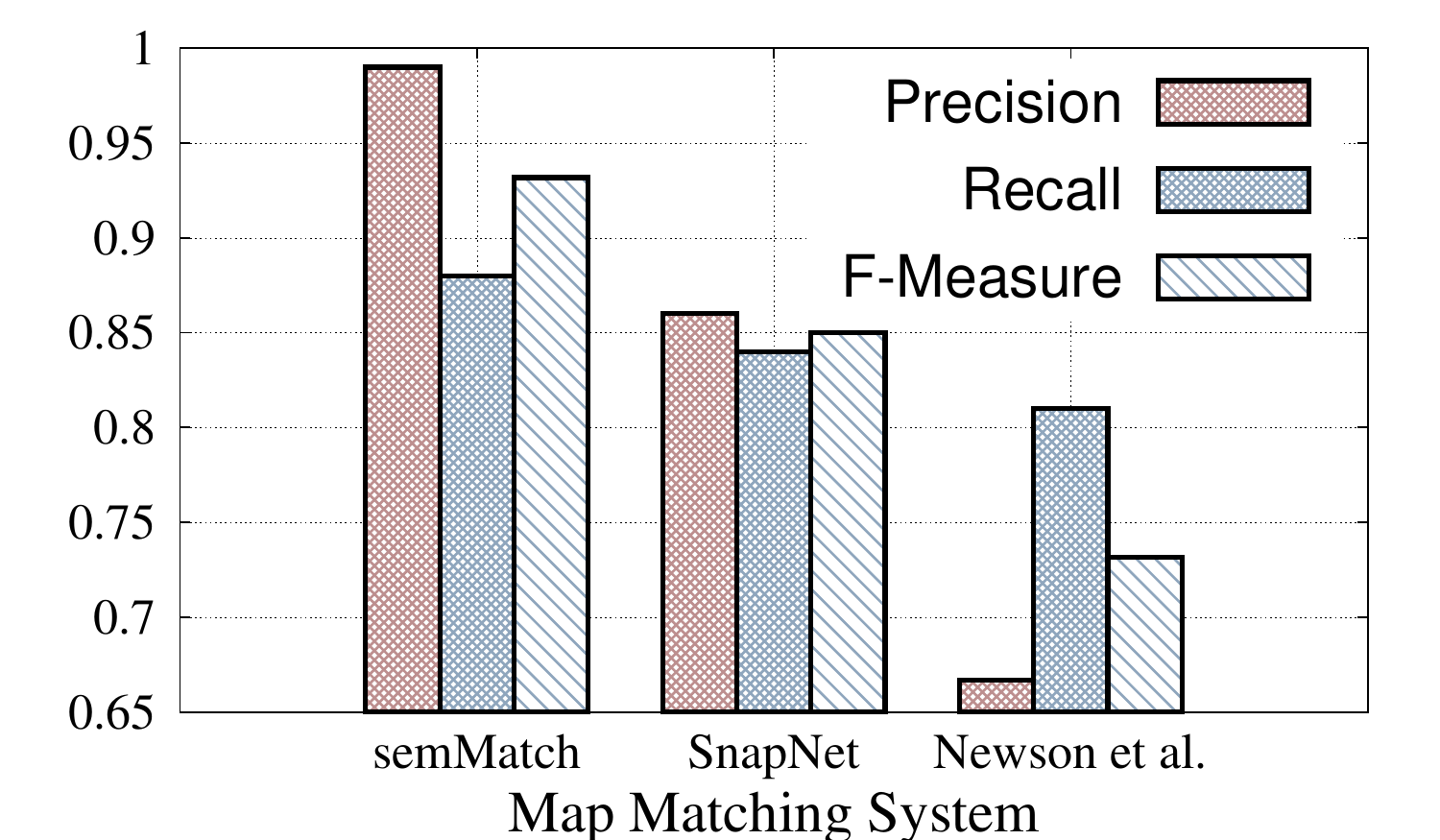}
      \label{fig:overall_cell}
    }\rulesep
 \subfigure[Network Pos. Data ``Cellular+WiFi'' (a large number of back-and-forth transitions)]{
      \includegraphics[width=0.3\linewidth]{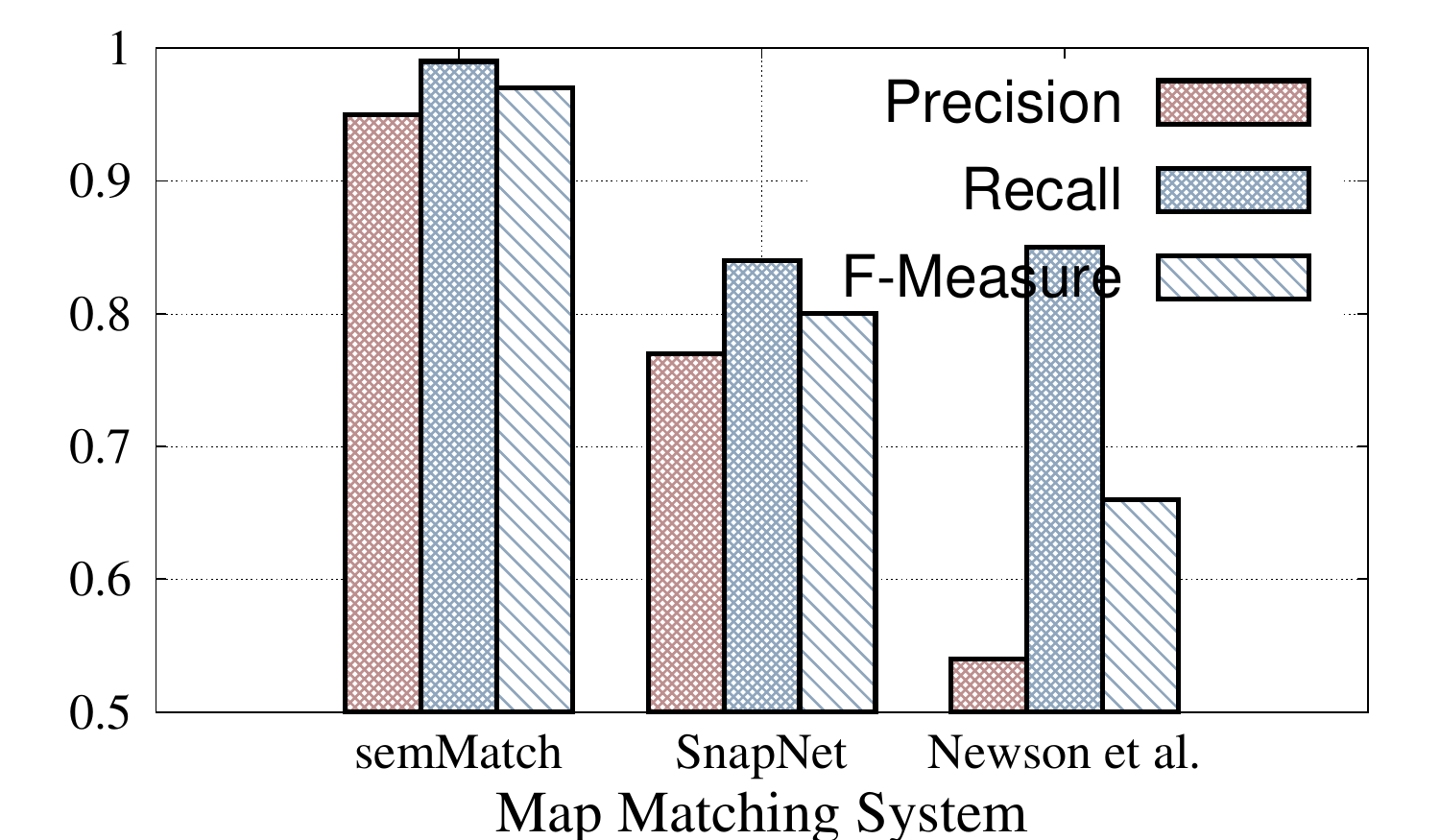}
      \label{fig:overall_nw}
    }\rulesep
 \subfigure[Low Sampling Rate GPS Positioning Data (1 sample every 2 mins)]{
      \includegraphics[width=0.3\linewidth]{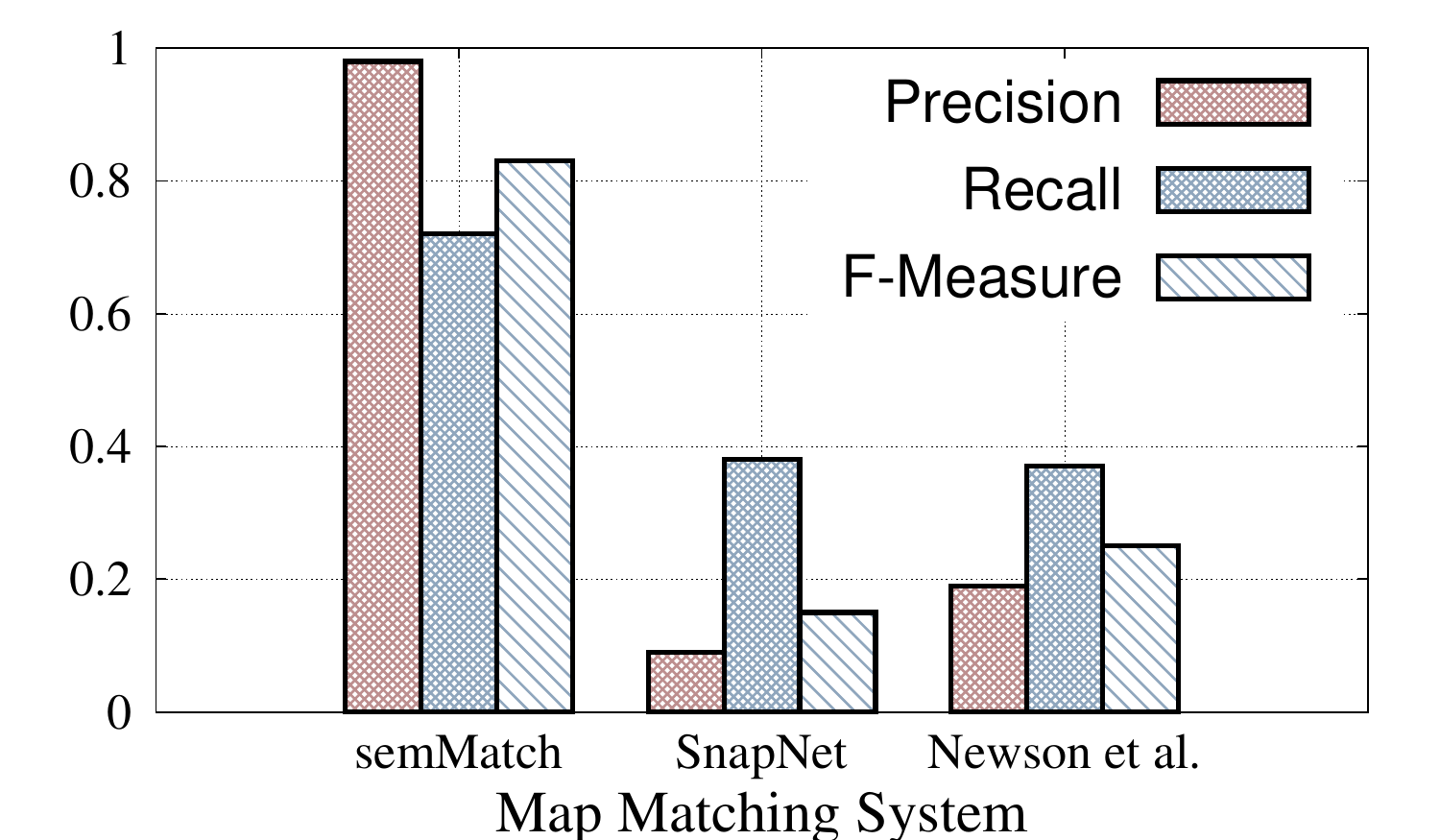}
      \label{fig:overall_gps}
    }
\caption{Performance of \sys{} as compared to SnapNet~\cite{mohamed2014accurate} and Newson et al~\cite{newson2009hidden} under different \textbf{challenging positioning data} classes.}
\label{fig:overall}
\vspace*{-5mm}
\end{figure*}
\section{Experimental Evaluation}\label{sec:eval}

In this section, we show our evaluation for the performance of \sys{}. We stress-tested the system using real challenging positioning data traces including: (1) Cellular-based traces with accuracy in the order of kilometers, (2) Network-based (Cellular+WiFi) with a large number of back-and-force transitions, and (3) Low-sampling rate GPS-based trajectories (one sample every two minutes).  Our data consists of over 150~km of car driving traces collected in two different cities. The data was collected using different Android phones including Samsung Galaxy S3 Mini, S5, and LG Nexus 4. We also used OpenStreetMaps for the road network information.

To get the cellular-based location estimates, we use the Google Android Location API \textbf{with WiFi turned off}. This way, the returned positioning estimates are based on the associated cell tower information only since the phones we used does not provide the neighboring cell towers information. The cellular-positioning data had \textit{an average accuracy of 1.9 km and density of 1.4 location estimate per km}. To get the network-based location estimates, we use the Google Android Location API \textbf{with WiFi turned on}.  The network-positioning data had \textit{an average accuracy of 162m and density of 6.8 location estimate per km}. For the low-sampling rate GPS-data, we used the cellphones' GPS sensor with a sampling rate of one sample every two minutes.  The GPS-positioning data had \textit{an average accuracy of 19m and density of 0.6 location estimate per km}. Cellular and network-based traces are noisier than GPS-based traces and has much higher errors. However, for energy efficiency, the GPS is sampled at low sampling rates; leading to extremely sparse positioning data.

For the ground truth, we collected GPS/GlONASS-based location data at high sampling rate using an external receiver. Then, we map matched the traces using offline HMM-based map matching. It has been shown in literature that GPS traces with high sampling rate (> 1 sample per second) can be map matched to reconstruct the driving path \cite{newson2009hidden,mohamed2014accurate}.

The rest of this section is organized as follows: We start by describing our evaluation metrics. After that, we show the effect of the road semantics density on \sys{}'s performance. Then, we quantify the \sys{} map matching performance as compared to two map matching techniques from literature. Finally, we show the power consumption overhead when using \sys{}.

\subsection{Evaluation Metrics}

To evaluate \sys{} map matching accuracy, we find the common matched road segment sequence between the output map matched trajectory and the ground-truth trajectory.
Then, we compute the map matching \emph{precision and recall} using this common sequence (Figure~\ref{fig:metrics}). The map matching precision shows how accurate the map matched trace is; it is defined as the ratio between the distance traversed on the matching sequence $X$ and the total distance of the map matched trajectory ($X+Y$). On the other hand,  the map matching recall shows the actual trajectory recovered ratio; it is defined as the ratio between the distance traversed on the matching sequence $X$ and the total distance of the ground truth trajectory $G$.

More formally, we can define the metrics as:
\begin{equation}
\begin{array}{ll}
 \textit{Precision} &= \frac{\text{Total distance of common matching sequence }}{\text{Total distance of output trace }}\\
 &= \frac{X}{X+Y}
 \end{array}
\end{equation}

\begin{equation}
\begin{array}{ll}
 \textit{Recall} &= \frac{\text{Total distance of common matching sequence }}{\text{Total distance of ground truth }}\\
 &= \frac{X}{G}
 \end{array}
\end{equation}

We also calculate the map matching F-Measure (F$_1$ score) which is the harmonic mean of the precision and recall:
\begin{equation}
 \textit{F-Measure} = 2\times\frac{\textit{Precision}\times \textit{Recall}}{\textit{Precision}+\textit{Recall}}
\end{equation}
\subsection{Effect of Road Semantics Density on Accuracy}

Figure~\ref{fig:semdensity} shows the density of the different road semantics encountered in our trajectories. On average, the density of the different road semantics is 2 semantics per km; this is higher than the cellular location update rate by 43\%, which allows \sys{} to achieve better map matching accuracy.

Figure~\ref{fig:density} shows the effect of the road semantics density on the map matching precision and recall. The figure shows the higher the density, the higher the system accuracy. \sys{} can achieve an F-measure of 93\% using the typical road semantics density. 
\subsection{Comparison with Other Systems}

In this section, we compare the performance of \sys{} in terms of precision, recall, and F-measure to two traditional HMM map matching algorithms: the one by Newson et al~\cite{newson2009hidden} and the SnapNet system~\cite{mohamed2014accurate}.
Newson et al~\cite{newson2009hidden} use the standard HMM-based algorithm to map match noisy and relatively sparse (up to one sample every 30~sec.) GPS traces. On the other hand, SnapNet targets map matching noisy cellular-positioning data. They apply several preprocessing stages based on the road network topology and assumed that the users favor major roads to overcome the imposed cellular-based challenges.

We used different scenarios of challenging positioning data: (1) Cellular-based data (avg. position accuracy=1.9 km, avg. location update= 1.4 location estimate per km), (2) Network-based data (WiFi+Cellular) (noisy data with a lot of back-and-forth transitions), and (3) GPS with low sampling rate (1 sample every 2 minutes).

Figure~\ref{fig:overall} shows that \sys{} has the \textbf{\emph{significantly better accuracy compared to the other two techniques under all scenarios}}. For the most challenging case of extremely low sampling rate of one location estimate every two minutes, \sys{] leads to enhancement in the (precision, recall, F-measure) of (989\%, 894\%, 232\%) over the Newson et al method~\cite{newson2009hidden} and (416\%, 946\%, 453\%) over the SnapNet system. This highlights the advantage of incorporating the road semantics in the map matching process under different harsh conditions.

\subsection{Power Consumption}
\begin{figure}[!t]
\centering
      \includegraphics[width=0.8\linewidth]{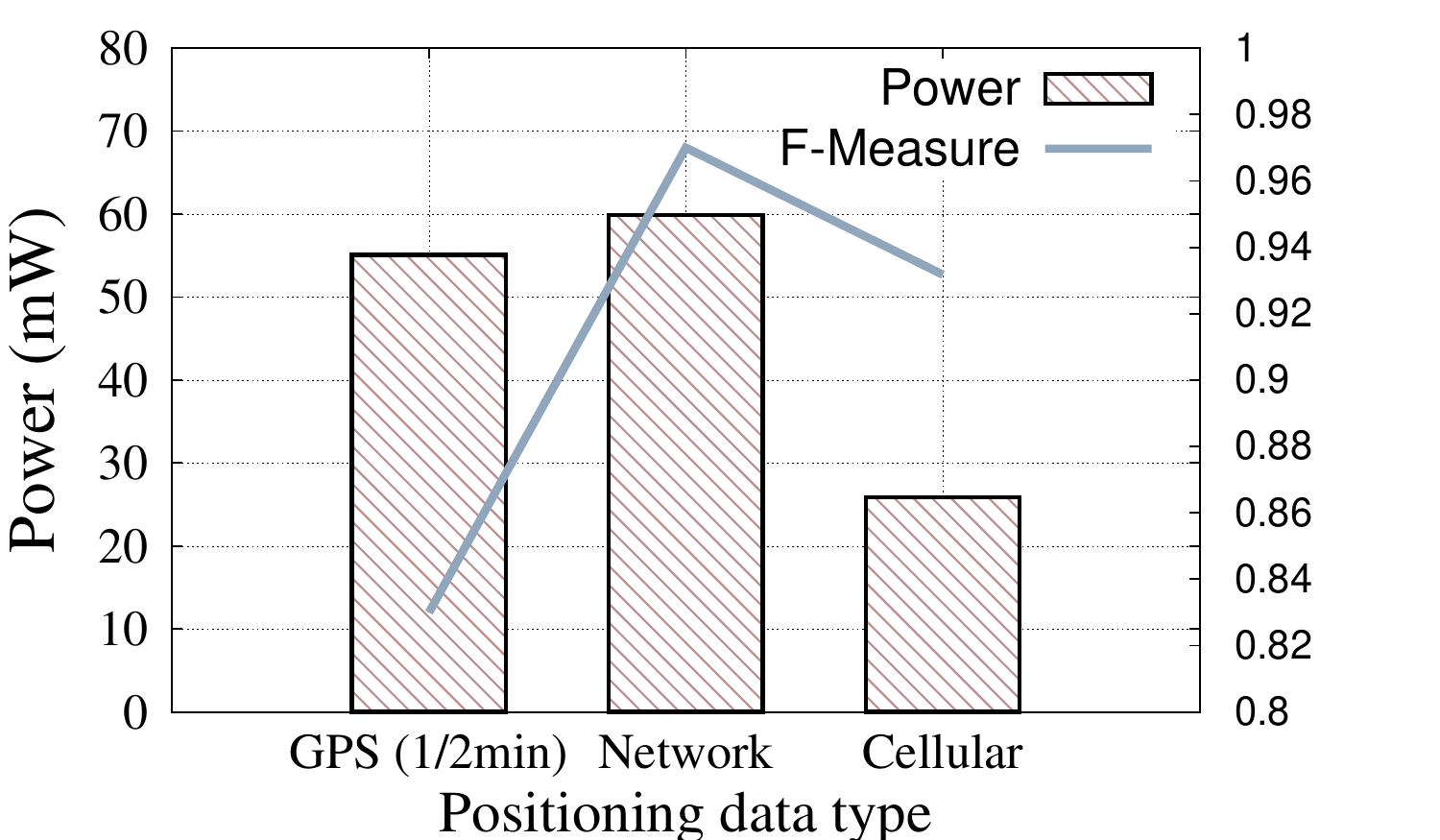}
\caption{\sys{} power consumption for the different positioning data types (low sampling rate GPS, network, and cellular) and the corresponding map matching F-measure.}
\label{fig:power}
\end{figure}
\sys{} collects inertial sensor measurements along with the positioning data to detect the different road semantics and improve the map matching accuracy. Figure~\ref{fig:power} shows the system's power consumption along with the map matching F-measure. The figure shows that \sys{} could achieve high map matching accuracy for the different challenging positioning data types with a low-energy footprint. Note that \sys{} depends on the smartphone's inertial sensors to detect the road semantics. Inertial sensors have low power consumption and are already running for other purposes (e.g. screen orientation detection). Hence, using them for road semantics detection consumes zero extra energy.

\section{Related Work}\label{sec:rw}
\subsection{Map Matching Techniques}
Due to its importance in many applications, map matching algorithms have gained due attention from researchers~\cite{greenfeld2002matching,quddus2003general,yang2005map,li2007practical,alt2003matching,brakatsoulas2005map,hummel2006map,newson2009hidden,mohamed2014accurate}.
These algorithms employ different techniques including geometric, topological, and probabilistic ones~\cite{quddus2007current}. Geometry-based map matching algorithms utilize the geometry and shape of the input trace and the road network. For example, in~\cite{white2000some}, authors map match the input location point to its closest arc (sequence of road map nodes). Similarly, in~\cite{white2000some,phuyal2001method}, authors use geometric curve to curve matching; where they map match a series of points simultaneously to the closest arc.  These techniques consider only the shape of the links regardless of their connectivity, leading to false transitions. Moreover, while they could provide good map matching accuracy for clean and dense GPS trajectories, they are unsuitable for noisy and sparse positioning data. 
On the other hand, topology-based map matching algorithms utilize the road segments topological information, e.g. connectivity and contiguity, along with the input trace and the road network shapes~\cite{quddus2003general,greenfeld2002matching,blazquez2005simple}. For example, in~\cite{quddus2003general}, authors use various similarity criteria (e.g.~vehicle speed, closeness) between the road network geometry and derived GPS navigation data to map match the user trajectory to the road network.  While topology-based algorithms improve upon geometry-based ones by considering road topology and characteristics, they are still vulnerable to the highly erroneous and sparse data.

Finally, probabilistic map matching algorithms employ probabilistic frameworks such as HMM or particle filters to map match the input user trajectories~\cite{hummel2006map,newson2009hidden,gustafsson2002particle}. For example, in \cite{hummel2006map,newson2009hidden}, HMM-based techniques give relatively good accuracy for the relatively noisy GPS trajectories with up to 30 seconds delay. However, highly noisy and sparse data are much harder to map match as shown in Figure~\ref{fig:cell_challenges}. Hence, their accuracy degrades significantly, up to 0.25 map matching F-Measure for low sampling rate GPS trajectories (as we quantified in Section~\ref{sec:eval}).

Recently, few systems were proposed to solve the map matching problem for coarse-grained network-based location data~\cite{vtrack,ctrack,autowitness,wheelloc,mohamed2014accurate}. Usually, these techniques try to enhance the accuracy of the input location data by leveraging other sensors. For example, VTrack \cite{vtrack} builds an HMM-based map matching scheme leveraging the WiFi data to handle the inaccuracy of cellular location information. CTrack \cite{ctrack} alternatively uses \emph{war-driving training} data in addition to inertial sensors to reduce the inaccuracy of cellular-based locations. Collecting war-driving data is an expensive and time-consuming process. Moreover, the cellular war-driven data typically has heterogeneity problems and need to be updated from-time-to-time. The AutoWitness~\cite{autowitness} and WheelLoc~\cite{wheelloc} systems leverage inertial sensors and dead-reckoning to reduce the inaccuracy of cellular-based localization. However, using the accelerometer and compass for dead-reckoning leads to fast error accumulation for their estimated trajectories~\cite{aly2013dejavu,mohssen2014s,uptime,SemSLAM}.
 SnapNet~\cite{mohamed2014accurate} targets cell-Id based positioning data by leveraging information from the digital map including road types and speed; they favor major roads and staying on the same road. Hence, the system performance would degrade when the user traverses side roads which are typically a part of any trip.

\sys{} is \textbf{\emph{unique in employing a general road sem- antics-based map matching framework}} that leverages a wide range of road semantics landmarks such as tunnels, curves and bridges to improve the map matching accuracy through a novel HMM-based model. \sys{} takes advantage of the inertial sensors \textbf{\emph{energy efficiency}} to detect the various road semantics and avoid their dead-reckoning issues. This leads to \textbf{\emph{enhancements in the map matching  F-Measure of at least $232\%$}} when compared to traditional and state-of-the-art HMM map matching algorithms~\cite{newson2009hidden,mohamed2014accurate} as quantified in Section~\ref{sec:eval}.

\subsection{Road Semantics Detection}
Inertial sensors have been used in literature for monitoring road conditions, e.g.~\cite{pothole,nericell,mednis2011real}. In~\cite{pothole,nericell}, external accelerometers were used to detect potholes and traffic conditions while in~\cite{mednis2011real}, the smartphone's accelerometer was used to detect potholes. All these systems use GPS to localize the road problems.
 In~\cite{aly_map14}, we extended these systems by leveraging different smartphone's sensors to detect a wide range of outdoor road semantics with the goal of automatically enriching the current digital maps. \sys{} builds on these road-semantics enriched digital-map to provide accurate map matching for challenging positioning data. 
\section{Conclusions}\label{sec:con}
We presented \sys{}: a novel road semantics-based  map matcher for the challenging cellular-based trajectories. \sys{} leverages the smartphone's inertial sensors to detect different road semantics and uses them in a mathematically-principled way to improve the accuracy and efficiency of the HMM map matching algorithm. We provided the \sys{}'s system architecture and our different preprocessing modules that help it reduce the noise in the input data. Moreover, we presented our novel incremental road semantics-based HMM algorithm in detail.

Evaluation of \sys{} on different classes of challenging positioning traces collected from two different cities covering  more than 150km shows that \sys{} can significantly outperform the traditional map matching algorithms in all scenarios reaching an F-Measure up to 97\%. This maps to at least 232\% enhancement in the F-measure in the most challenging case. In addition, \sys{} has a low energy footprint on the scarce phone battery.

\section*{Acknowledgment}
This work is supported in part by a grant from the Egyptian Information Technology Industry Development Agency (ITIDA). We also sincerely thank \textit{Reham Mohamed} for her valuable contributions and suggestions to the \sys{} algorithm.

\bibliographystyle{abbrv}
{
\bibliography{ms}
}
\end{document}